\documentclass[10pt,aps,prb,nobibnotes,superscriptaddress,secnumarabic,eqsecnum,twocolumn,floatfix]{revtex4-1}
\usepackage{amssymb}
\usepackage{eurosym}
\usepackage{color}
\usepackage[utf8]{inputenc}
\usepackage{graphicx}
\usepackage{graphics}

\UseRawInputEncoding
\def\ba{\begin{eqnarray}}
\def\ea{\end{eqnarray}}
\def\be{\begin{equation}}
\def\ee{\end{equation}}

\begin{document}

\title{Solitonic self-sustained charge and energy transport on the
superconducting cylinder}
\author{\vspace{2mm} Fabrizio Canfora}
\email{fabrizio.canfora@uss.cl}
\address{
\vspace{1.8mm}
{Universidad San Sebasti\'an, sede Valdivia, General Lagos 1163, Valdivia 5110693, Chile}\\}
\address{
\vspace{1.8mm}
{Centro de Estudios Cient\'ificos (CECS) Casilla 1469, Valdivia, Chile}}
\author{Alex Giacomini}
\email{alexgiacomini@uach.cl}
\address{
\vspace{1.8mm} 
{Instituto de Ciencias F\'isicas y Matem\'aticas, Universidad Austral de Chile, Casilla 567, Valdivia, Chile}}
\author{Nicol\'as Grandi}
\email{grandi@fisica.unlp.edu.ar}
\address{
\vspace{1.8mm} 
{Departamento de F\'isica, UNLP, C.C. 67, 1900 La Plata, Argentina}\\}
\address{
\vspace{1.8mm} 
{Instituto de F\'isica de La Plata, CONICET, C.C. 67, 1900 La Plata, Argentina}}
\author{Julio Oliva}
\email{juoliva@udec.cl}
\address{
{Departamento de F\'isica, Universidad de Concepci\'on, Casilla 160-C, Concepci\'on, Chile} 
}
\author{Aldo Vera$^3,$}
\email{aldo.vera@uach.cl}

\begin{abstract}
\vspace{0.3mm}
 We present an exact time-dependent solution for a charged
scalar field on a two-dimensional cylinder, that can be interpreted as
representing a long-standing excitation on a $s$-wave superconducting state,
which propagates along a nanotube constructed out of twisted bilayer
graphene. The solution has a topological charge characterized by an integer number,
which counts the winding of the Higgs phase winds around the cylinder. The resulting
electric current generates its own electromagnetic
field in a self-consistent way, without the need of any external fields to
keep it alive.

\vspace{1.2mm}
\end{abstract}

\maketitle

\section{Introduction}

\label{sec:intro}

The charged scalar field in two space dimensions is a very simple yet very
useful quantum field theory. It has been studied as a lower dimensional toy
model for quantum electrodynamics\cite{algunlibro}, as a constructive
laboratory to investigate the properties of solitonic solutions in gauge
field theories \cite{elindioese} and, most importantly, in the condensed
matter realm, as the effective Ginzburg-Landau theory describing a
two-dimensional $s$-wave superconductor\cite{ginzburglandau}.

In the last context, the charged scalar represents the degrees of freedom of
the superconducting condensate. The homogeneous solution with a
non-vanishing scalar expectation value is interpreted as the superconducting
state, since it spontaneously breaks the $U(1)$ charge invariance. Static
inhomogeneous solutions then represent excitations; a well known example of
which is the Nilsen-Olesen vortex\cite{nilsenolesen}, which depicts an
Abrikosov vortex on the superconducting background \cite{abrikosov}.

Understanding the dynamics of such excitations requires the inclusion of
time derivatives in the scalar field equations of motion. First order time
derivatives result in the time-dependent Ginzbug-Landau theory\cite%
{tdginzburglandau1, tdginzburglandau2}. This model is non-invariant under
time reversal and then it results in dissipation. Alternatively, a Lorentz
covariant approach has been suggested, in which the dynamics corresponds to $%
(2+1)$-dimensional scalar electrodynamics\cite{holandos}. It has the
advantage of being consistent with the covariance of Maxwell equations, as
well as providing a finite penetration depth for the electric field\cite%
{hirsch}. Dissipation due to non-condensed electrons can then be included
through a Rayleigh dissipation function \cite{grigorishin}.

Recently, unconventional high T$_c$ superconductivity has been observed in
twisted bilayer graphene\cite{nature}, when the twist is adjusted at the so
called ``magic angle'' $\sim 1.1^\circ$ at which the fermionic dispersion
relation develops a flat band\cite{flatband1, flatband2}. It is not clear
yet what is the symmetry of the resulting wave function. It has been argued
that single\cite{swave1} and multilayer\cite{swave2} graphene might form $s$%
-wave pairs, of which some experimental evidence has been found\cite{swave3}.

In the present paper, inspired in the above considerations, we explore the
solutions to $(2+1)$-dimensional scalar electrodynamics. We are interested
in long-standing time-dependent configurations. Hence, the Ansatz must be
chosen in such a way as to minimize all possible sources of dissipation. We
find a traveling non-linear wave that moves along the non-compact direction
of a cylinder, at the speed of light. We interpret it as an excitation
propagating in the superconducting state of a nanotube constructed out of
twisted bilayer graphene.

A particularly interesting problem is whether the energy and charge
transport can be self-sustained. In other words: {is an external field
necessary to keep these excitations alive, or is it enough to consider the
electromagnetic field generated by the excited charges themselves?} Such a
question is notoriously difficult (see \cite{self1, self2, self3} and
references therein) since it entails to consider the back-reaction of the
excitations on the electromagnetic field and viceversa. Our construction
shows that, at least in the present setting, the second possibility is a
viable option.

The paper is organized as follows: In Section \ref{sec:model} we define our
model and parametrize the fields. In Section \ref{sec:decoupling} we
investigate the decoupling conditions which minimize dissipation. In Section \ref{sec:solution} we write the explicit solution
and sections \ref{sec:transport} and \ref{sec:flux} are dedicated to the study of its transport and topological
properties. In Section \ref{sec:examples} we discuss some specific examples. In Section \ref{sec:stability},
we analyze the stability of a special type of perturbations. Finally, in Section \ref{sec:discussion} we discuss our
findings.

\section{The model}
\label{sec:model}

The effective degrees of freedom of a relativistic $s$-wave superconductor
are described in a Lorentz covariant way by standard scalar electrodynamics
in $2+1$ dimensions\cite{holandos, grigorishin}, which couples a charged
scalar field $\Psi$ to the electromagnetic field $A_\mu$ minimally,
according to the action 
\begin{widetext}
\be
S=-\frac 12\int_{\Omega\times\mathbb{R}}  d^3x  \sqrt{-g}\left((D_\mu\Psi)^* D^\mu\Psi+\frac\gamma2 \left(|\Psi|^2-\nu^2\right)^{2}+\frac12 F_{\mu\nu}F^{\mu\nu}\!\right)\,,
\ee
\end{widetext}
where the system is defined in a spatial manifold $\Omega$, being $D_\mu=\partial_\mu-ieA_\mu$ the covariant derivative, $e$ the
scalar electric charge, $\gamma$ its coupling constant, and $\nu$  its vacuum expectation value.

We parametrize the scalar in the most general way, as
\begin{equation}
\Psi=he^{i eG}\,,  \label{scalar}
\end{equation}
where the factor of $e$ in the exponent is introduced for later convenience.
Regarding the gauge field, we use a ``Clebsch representation'' of the form 
\begin{eqnarray}
A_\mu =\partial_\mu\Lambda+\lambda\,\partial_\mu F\,,  \label{vector}
\end{eqnarray}
in terms of the ``Clebsch potentials'' $\Lambda$, $\lambda$ and $F$. In $2+1$
dimensions this decomposition is completely general and does not restrict the
fields in any sense. Of course the gradient part $\partial_\mu\Lambda$ can
be adjusted to any desired value by a gauge transformation, but we leave
that for later after deriving the equations of motion.

Since the parametrization is not restrictive, we can safely plug it into the
action and then obtain the equations of motion by varying $h,
G,\Lambda,\lambda$ and $F$. To do that, we need the expressions for the
gauge curvature and the covariant derivative 
\begin{eqnarray}
&&\!\!\!\!\!\!\!\!F_{\mu \nu }=\partial _{\mu }\lambda \partial _{\nu
}F-\partial _{\nu }\lambda \partial _{\mu }F\,, \\
&&\!\!\!\!\!\!\!\!D_{\mu }\Psi =\left( \partial _{\mu }h+ieh(\partial _{\mu
}G-\partial _{\mu }\Lambda -\lambda \partial _{\mu }F)\right) e^{ieG}\,,
\end{eqnarray}
which, together with Eqs. (\ref{scalar}) and (\ref{vector}), result in the
rewritten form of the action 
\begin{widetext}
\be
S=\frac12 \int_{\Omega\times\mathbb{R}} d^3x \sqrt{-g}\left(-(\partial h)^2 -{e^2}(\partial G-\partial\Lambda-\lambda \partial F)^2h^2-\frac \gamma2 \left(h^2-\nu^2\right)^2+(\partial\lambda\cdot\partial F)^2-(\partial\lambda)^2(\partial F)^2\right)\,.
\label{actionn}
\ee
\end{widetext}
This can be varied with respect to the different fields, to obtain the
equations of motion of the system. We start with the scalar phase $G$,
obtaining the equation 
\begin{equation}
\partial ^{\mu }\left( (\partial _{\mu }G-\partial _{\mu }\Lambda -\lambda
\partial _{\mu }F)h^{2}\right) =0\,.  \label{qlo}
\end{equation}
Notice that the same equation can be obtained by varying with respect to $%
\Lambda $. Then, we can choose the gauge $\Lambda =G$ and, replacing it into
the action, the remaining equations take the form 
\begin{eqnarray}
&&\!\!\!\!\!\!\!\!\!\!\!\!\!\!\!\!\partial ^{\mu }\!\left( (\partial
F)^{2}\partial _{\mu }\lambda -(\partial \lambda \cdot \partial F)\partial
_{\mu }F\right) -{e^{2}}\lambda h^{2}(\partial F)^{2}=0\,,  \label{qqlo1} \\
&&\!\!\!\!\!\!\!\!\!\!\!\!\!\!\!\!\Box h-e^{2}h\lambda ^{2}(\partial
F)^{2}-\gamma (h^{2}-\nu ^{2})h=0\,,  \label{qqlo2} \\
&&\!\!\!\!\!\!\!\!\!\!\!\!\!\!\!\!\partial ^{\mu }\left( (\partial \lambda
\cdot \partial F)\partial _{\mu }\lambda -\left( (\partial \lambda )^{2} +{%
e^{2}}h^{2}\lambda ^{2}\right) \partial _{\mu }F\right) =0\,,  \label{qqlo3}
\end{eqnarray}%
while Eq. (\ref{qlo}) simplifies to 
\begin{equation}
\partial ^{\mu }\!\!\left( h^{2}\lambda \partial _{\mu }F\right) =0\,.
\label{qqqlo}
\end{equation}%
Equations (\ref{qqlo1})-(\ref{qqqlo}) constitute the full set of equations
of motion of the system.

\section{Long term behavior}
\label{sec:decoupling}

In the above equations (\ref{qqlo1})-(\ref{qqqlo}) the electromagnetic
variables $F$ and $\lambda $ are coupled among them and to the scalar one $h$.
This means that any amount of energy stored on any of the fields will get
distributed among all the degrees of freedom as the system evolves. In any
realistic situation, each variable is coupled to an external dissipative
channel. This implies that the energy will then flow out of the system
through all of them. If one waits long enough, it is reasonable to expect
that one should be left with a configuration in which all the possible
dissipative processes have already happened. Thus, the Ansatz describing
such configuration must minimize the coupling among the different
dissipative channels: this happens when the different degrees of freedom of
the gauge and scalar fields are decoupled. In that case, the energy stored
on any degree of freedom dissipates only through the corresponding channel,
resulting in less dissipation overall.

These intuitive arguments lead to the following consistent Ansatz. First of
all, one needs to impose the \textquotedblleft force free\textquotedblright\
conditions $\partial \lambda \cdot \partial F=\partial h\cdot \partial F=0$
and the \textquotedblleft light-like\textquotedblright\ condition $(\partial
F)^{2}=0$, since in this way two degrees of freedom (the gauge field and the
amplitude of the scalar field) get decoupled. Then, the additional condition 
$\partial ((\partial \lambda )^{2})\cdot \partial F=0$ must be imposed for
consistency. In this way, the full set of coupled field equations is reduced
to the decoupled pair
\begin{eqnarray}
&&\square h-\gamma (h^{2}-\nu ^{2})h=0\,, \\
&&\square F=0\,,
\end{eqnarray}
where we assumed $(\partial \lambda )^{2}+{e^{2}}h^{2}\lambda ^{2}\neq 0$.
These equations must be solved together with the constraints 
\begin{eqnarray}
&&(\partial F)^{2}=0\,,\qquad \qquad \label{uno}\\
&&\partial \lambda \cdot \partial F=0\,,\;\quad \qquad \label{dos}\\
&&\partial h\cdot
\partial F=0\,,\\
&&\partial ((\partial \lambda )^{2})\cdot
\partial F=0\,,
\end{eqnarray}
to obtain a full solution of the system.
\section{The superconducting cylinder}
\label{sec:solution}

The next step is to specify the space-time geometry $\Omega\times \mathbb{R}$ in which our fields propagate. We
choose a cylinder topology with planar metric 
\begin{equation}
ds^{2}=-dt^{2}+dz^{2}+d\varphi ^{2}\,,
\end{equation}%
where the coordinate $\varphi \approx \varphi +L_\varphi$ goes around the
cylinder, while $-{L_{z}}/{2}<z<{L_{z}}/{2}$ runs along it (the case of a
very long cylinder corresponds to $L_{z}\rightarrow +\infty $). We can
change to lightlike coordinates $z_{\pm }=(z\pm t)/\sqrt{2}$ obtaining 
\begin{equation}
ds^{2}=2dz_{+}dz_{-}+d\varphi ^{2}\,,
\end{equation}%
which implies in the equations of motion 
\begin{eqnarray}
&&2\partial _{+}\partial _{-}h+\partial _{\varphi }^{2}h -\gamma (h^{2}-\nu
^{2})h=0\,, \\
&&2\partial _{+}\partial _{-}F+\partial _{\varphi }^{2}F=0\,,
\end{eqnarray}%
and in the constraints 
\begin{eqnarray}
&&\!\!\!\!\!\!\!\!2(\partial _{+}F)(\partial _{-}F)+(\partial _{\varphi
}F)^{2}=0\,, \\
&&\!\!\!\!\!\!\!\!\partial _{+}\lambda \,\partial _{-}F+\partial _{-}\lambda
\,\partial _{+}F+\partial _{\varphi }\lambda \,\partial _{\varphi }F=0\,, \\
&&\!\!\!\!\!\!\!\!\partial _{+}h\,\partial _{-}F+\partial _{-}h\,\partial
_{+}F+\partial _{\varphi }h\,\partial _{\varphi }F=0\,, \\
&&\!\!\!\!\!\!\!\!\partial _{+}((\partial \lambda )^{2})(\partial
_{-}F)\!+\!\partial _{-}((\partial \lambda )^{2})(\partial
_{+}F)\!+\!\partial _{\varphi }((\partial \lambda )^{2})\partial _{\varphi
}F=0\,.  \nonumber \\
&&
\end{eqnarray}%
The lightlike condition in the first line can then be solved by first
choosing $\partial _{\varphi }F=0$ and then imposing $\partial _{+}F=0$ or $%
\partial _{-}F=0$. The force free conditions in the second and third lines
then imply that $\lambda $ and $h$ should depend on the same $z_{\pm }$
variable as $F$. With these choices, the condition on the last line as well
as the equation of motion for $F$ are automatically satisfied. We are then
left with only one equation of motion, the one for $h$, which now reads 
\begin{equation}
\partial _{\varphi }^{2}h-\gamma (h^{2}-\nu ^{2})h=0 \ .  \label{QuOsc}
\end{equation}%
This equation corresponds to a quartic oscillator, and can be integrated once
to get the first order relation 
\begin{equation}
\frac{\left( \partial _{\varphi }h\right) ^{2}}{2}-\frac{\gamma }{4}\left( {%
h^{2}}-\nu ^{2}\right)^2 =-\frac {\gamma \nu^4}4 \left(\frac{m-1}{m+1}%
\right)^2 \ ,  \label{minustphiphi}
\end{equation}
where the integration constant in the right hand side was conveniently parametrized in
terms of a new number $m$. This can reduced to quadratures, as
\begin{equation}
\int\frac{dh}{\sqrt{\left( {h^{2}}-\nu ^{2}\right)^2 - \nu^4\left(\frac{m-1}{%
m+1}\right)^2}}=\pm\sqrt{\frac{\gamma }{2}}(\varphi -\varphi_0) \ ,
\label{QuOsc1}
\end{equation}%
where $\varphi_0$ is a new integration constant. 
The remaining integral can be performed explicitly. 

For the particular case $m=1$ we get the solution
\begin{equation}
h=\nu\tanh\left(\nu\sqrt{\frac\gamma2}(\varphi-\varphi_0)\right) \ ,
\label{m=1}
\end{equation}
where an overall $\pm$ sign was removed by a gauge transformation. Is is evident that this function approaches
the vacuum value at $\varphi\to\pm\infty$. This implies that the natural boundary condition, namely that the
Higgs field matches its vacuum value at the boundary of the sample, can only be satisfied in an infinite plane.

If $m\neq1$ the solution can be written in terms of the Jacobi elliptic sine function, as 
\begin{equation}
h=\nu \sqrt{\frac{2m}{1+m}}\,\mathrm{sn}\!\left( \nu \sqrt{\frac{\gamma }{1+m%
}}\,(\varphi -\varphi _{0}),m\right) \ ,  \label{eq:solutionhiggsraw}
\end{equation}
where again an overall $\pm$ sign was gauged away. In this expression, in order to have
a real $h$ we need a positive value for $m$.
Since the solution is invariant when $m\to1/m$, we only need to consider $%
0<m<1$. Here and accordingly to the constraint $\partial _{+}h=0$ or $%
\partial _{-}h=0$, the integration constants $m$ and $\varphi _{0}$ are
arbitrary functions of $z_{-}$ or $z_{+}$, respectively. These will be
restricted further as we impose physically meaningful boundary conditions on
the obtained solution.

\begin{figure}[t]
\includegraphics[scale=.5]{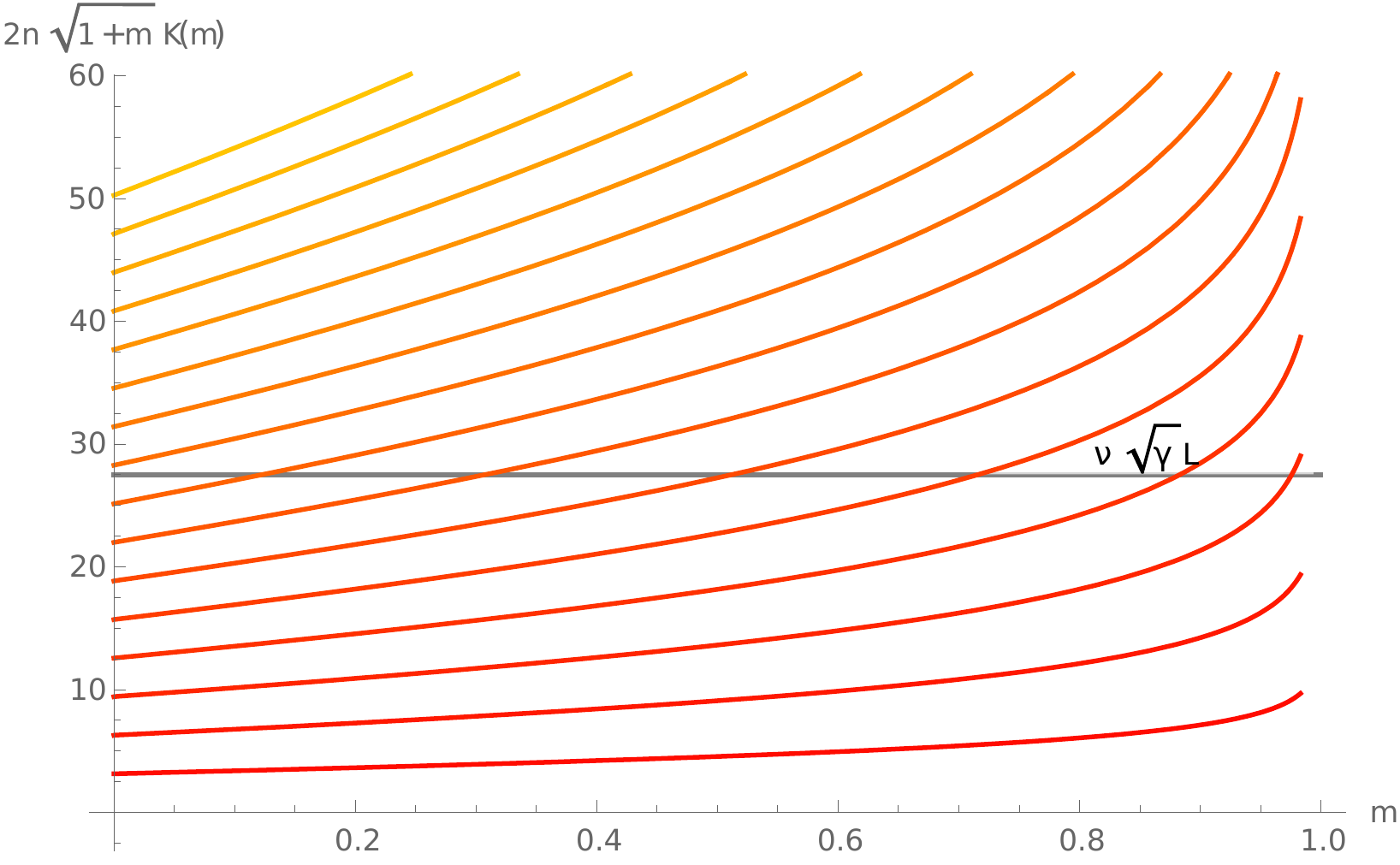}
\caption{From bottom to top, the profiles of the expression $4n\protect\sqrt{%
(1+m)}K(m)$ as a function of $m$ for growing $n=1,2,\dots,7$. The horizontal
line represents an arbitrary value of the combination $\protect\nu\protect%
\sqrt{{\protect\gamma}}\,L$ of the model parameters. This shows that the
equation $\protect\nu\protect\sqrt{{\protect\gamma}}\,L=2n\protect\sqrt{(1+m)%
}K(m)$ has a non-trivial solution for $\protect\nu\protect\sqrt{{\protect\gamma}}\,L> 
\protect\pi n $.}
\label{figura}
\end{figure}

If the solution is defined on a tube, then the value of $h(\varphi )$ must
differ from the value $h(\varphi +L_{\varphi })$ by a gauge transformation, 
\emph{i.e.} by a phase. Since $h$ is real, the only possible phase is an
overall sign $\pm $, implying a periodic or antiperiodic solution. This
imposes 
\begin{equation}
\nu \sqrt{{\gamma }}\,L_{\varphi }=2n\sqrt{(1+m)}K(m)\ ,  \label{nquantcond}
\end{equation}
where $n\in \mathbb{Z}$ is an arbitrary integer and $K(m)$ is the complete
elliptic integral of the first kind. Then $m$ is fixed to a constant value
independent of $z_{\pm }$. For the model parameters satisfying $\nu \sqrt{%
\gamma }\,L_{\varphi }>\pi n$, this equation has a non-trivial solution; see
Fig.\thinspace \ref{figura}. The solution is periodic for $n$ even and
antiperiodic for $n$ odd.

If instead we want to define the solution on a ribbon extending from $\varphi _{0}$ to $\varphi _{0}+L_{\varphi }$,
we need to impose that the Higgs field reaches its vacuum expectation value at the edges $h(\varphi_{0})=\pm h(\varphi _{0}+L_{\varphi })=\nu $.
It has to do it smoothly, thus we also need to satisfy $h^{\prime }(\varphi _{0})=\pm h^{\prime }(\varphi_{0}+L_{\varphi })=0$.
This is just a particular case of the (anti)periodicity conditions discussed in the previous paragraph, and imposes the same quantization
condition for $m$. However, since the overall factor satisfies $\sqrt{2m/(1+m)}<1$ for the allowed range of $m$, the value of $h$ is never
$\pm\nu$, implying that the solution cannot exist on a ribbon.

Notice that the solution (\ref{eq:solutionhiggsraw}) changes sign $n$ times as the variable $\varphi$ goes around the tube. This can be made explicit by writing
\begin{eqnarray}
h=\left\vert h\right\vert e^{i \left[\frac {n\pi\varphi}{L_\varphi}\right]}\,,
\end{eqnarray}
where $[\cdots]$ stands for the integer part. The exponent is then a discontinuous function, which jumps as $\varphi$ grows. However,
a smooth overall phase for the Higgs field $\Psi$ can be obtained by choosing the function $G$ in our Ansatz (\ref{scalar}) in the form
\begin{equation}
eG+\left[\frac {n\pi\varphi}{L_\varphi}\right]=\frac {n\pi\varphi}{L_\varphi}\,.
\end{equation}
With this form of $G$, the number $n$ is measuring the winding of the Higgs phase around the cylinder. 

\bigskip

The full solution of the system then reads 
\begin{eqnarray}
&&\!\!\!\!\!\!\!\!\!\!\!\!\!\!\!\!\!\!\!F=F(z_\pm)\,, \\
&&\!\!\!\!\!\!\!\!\!\!\!\!\!\!\!\!\!\!\!\lambda=\lambda(z_\pm ,\varphi)\,, \\
&&\!\!\!\!\!\!\!\!\!\!\!\!\!\!\!\!\!\!\!h=\nu \sqrt{\frac{2m}{1+m}}\,\mathrm{%
sn}\!\left( \nu \sqrt{\frac{\gamma }{1+m}}\,(\varphi -\varphi _{0}),m\right) \ , \\
&&
\!\!\!\!\!\!\!\!\!\!\!\!\!\!\!\!\!\!\!eG=\frac {n\pi\varphi}{L_\varphi}-\left[\frac {n\pi\varphi}{L_\varphi}\right]=e\Lambda
\,,
\end{eqnarray}
where $m$ is quantized as in Fig. \ref{figura} with $n\in\mathbb{Z}$, and the functions $F(z_\pm )$, $\varphi_0(z_\pm )$ and
$\lambda(z_\pm,\varphi)$ are chiral but otherwise completely arbitrary, being determined by the initial conditions.

\section{Charge and energy transport}
\label{sec:transport}
The electric current of the above solution can be written as twice the
coefficient of the gauge potential $\lambda\partial_\pm F$ in the action,
and it takes the form 
\begin{eqnarray}
&&J_\pm={-}e^2h^2\lambda \partial_\pm F\,, \\
&&J_\varphi=0\, ,
\end{eqnarray}
implying $J_t=\pm J_z=\pm J_\pm$. Here and along this section, the top
(respectively bottom) signs represent the solution that depends on $z_+$
(respectively $z_-$). The above result implies in particular that, in order to
have a well defined current, the function $\lambda(z_\pm,\varphi)$ has to be
periodic in the variable $\varphi$.

\bigskip

We can also calculate the electromagnetic field strength, which read 
\begin{eqnarray}
&&F_{\varphi \pm}=\partial_\varphi\lambda\,\partial_\pm F={-}%
\partial_\varphi(J_z/e^2h^2)\,, \\
&&F_{\pm\mp}=F_{\pm\pm}=0\,,
\end{eqnarray}
or in other words 
\begin{equation}  \label{EyB}
E_z=0 \ , \quad -B=\pm E_\varphi \equiv F_{\varphi\pm} \ .
\end{equation}
This results on a vanishing total magnetic flux $\Phi_B$ on the cylinder, as the integral
\begin{eqnarray}
\Phi_B&=&\int_\Omega B=\int d\varphi\,dz\,\partial_\varphi \left(\frac {J_z}{e^2h^2}%
\right)=
\nonumber\\&=&
\int dz\,\left.\frac {J_z}{e^2h^2}%
\right|_{\varphi}^{\varphi+L_\varphi}=0\,,
\end{eqnarray}
vanishes in virtue of the periodicity of $J_z/e^2h^2$.

\bigskip

Effective electric conductivities can be defined as quotients of the electric current components divided by the electric
field ones. This results in an infinite effective direct conductivity $\sigma_{zz}=J_z/E_z$, and an effective Hall conductivity with value
\begin{equation}
\sigma_{z\varphi}=J_z/E_\varphi={\mp} e^2 h^2/\partial_\varphi\log\lambda\,.
\end{equation}

\bigskip

Finally, the components of energy momentum tensor read 
\begin{eqnarray}
&&T_{\pm\pm}=-e^2h^2\lambda^2(\partial_\pm F)^2- (\partial_\pm
h)^2-F_{\pm\varphi}^2\,, \\
&&T_{\pm\mp}=\frac12(\partial_\varphi h)^2+\frac\gamma 4(h^2-\nu^2)^2\,, \\
&&T_{\varphi\varphi}=-\frac12(\partial_\varphi h)^2+\frac\gamma
4(h^2-\nu^2)^2\,, \\
&&T_{\varphi\pm}=-\partial_\varphi h\,\partial_\pm h\,,
\end{eqnarray}
which results on an energy density $T_{tt}=T_{\pm\pm}-T_{\pm\mp}$, %
implying that for a finite tube of length $L_z$ 
the total energy is finite. Notice that $T_{\varphi\varphi}$ is minus the
constant (\ref{minustphiphi}), while $T_{\pm\mp}$ is minus the Lagrangian in (\ref{actionn}) evaluated
on the restrictions (\ref{uno})-(\ref{dos}) with $G=\Lambda$.

\bigskip

It is important to notice that  
the field strengths, the electric current, and the energy momentum tensor,
are all periodic in $\varphi$, even in the case when $h$ is anti-periodic.

\newpage
\section{Topological charge and a BPS-like bound}
\label{sec:flux}
Since our solutions are characterized by an integer $n$ which is a winding number, we may wonder whether it
is related to some topological charge. If we write the standard form of the topological charge for $2+1$ scalar electrodynamics as
\begin{equation}
Q=i\int_\Omega d\Psi\wedge d\Psi^*=i\int_{\partial\Omega}\Psi\wedge d\Psi^* \ , 
\end{equation}
where $\Omega$ is now our cylinder and $\partial\Omega$ are the two circles at the cylinder ends, we get the explicit expression
\begin{eqnarray}
Q&=&i\int d\varphi \,\left(
\left.\Psi\partial_\varphi\Psi^*\right|_{z=\frac {L_z}2}
+
\left.\Psi\partial_\varphi\Psi^*\right|_{z=-\frac {L_z}2}
\right)
\nonumber\\
&=& 2n\pi \ . 
\end{eqnarray}
Then we see that our solutions are topological in nature, being characterized by a topological charge. 

An interesting point is that the charge $Q$ was obtained from the same expression that would result in the topological
charge of an Abrikosov-Nielsen-Olesen vortex. Thus, our solutions may in principle be continuously deformed into such
a vortex, without changing the value of $Q$.

\bigskip

In static configurations, the presence of a topological charge is often related to the existence of a Bogomol'nyi-Prasad-Sommerfield
(BPS) bound on the energy. A natural question is whether something similar may exist for the present time dependent solutions.
To check that, we evaluate the on-shell action of the configuration
\begin{equation}
S_{\sf on-shell}=-\int dz_+dz_-d\varphi \left( \frac{1}{2}(\partial _{\varphi }h)^{2}+%
\frac{\gamma }{4}(h^{2}-\nu ^{2})^{2}\right) \ ,
\end{equation}
which can be rewritten as
\begin{equation}
S_{\sf on-shell}\!=\!-\frac{1}{2}\!\int\! dz_+dz_-d\varphi \left( \partial _{\varphi }h-s%
\sqrt{\frac{\gamma }{2}}(h^{2}\!-\nu ^{2})\right) ^{2}\!+P ,
\label{BPS1}
\end{equation}
where $s=\pm1$ is a sign, and we have defined the magnitude $P$ according to 
\begin{equation}
P=-s\sqrt{\frac{\gamma }{2}}\int dz\,d\varphi \,(h^{2}-\nu
^{2})\,\partial _{\varphi }h\ .
\end{equation}
In this expression, the integral in $\varphi $ can be explicitly performed, resulting in
\begin{equation}
P=-s\sqrt{\frac{\gamma }{2}}\int dz\,\left. \left( \frac{h^{2}}{3}-\nu
^{2}\right) h\right\vert _{\varphi }^{\varphi +L_{\varphi }} \ , 
\end{equation}
this vanishes for periodic ($n$ even) solutions, but not for antiperiodic ($n $ odd) ones. In this last case, the result has the somewhat
disappointing feature of being dependent on the value of $\varphi $ where the circle is closed $\varphi \approx \varphi +L_{\varphi }$,
which difficults a proper physical interpretation of $P$ as a topological charge. However, we can write the following BPS-like bound 
\begin{equation}
S_{\sf on-shell}\leq P\ ,
\label{BPS}
\end{equation}
whose saturation implies the field equations, as expected from a standard BPS bound. Indeed, the above bound is
saturated when equation (\ref{minustphiphi}) is satisfied, for the particular case $m=1$. As we mention earlier, the corresponding
solution (\ref{m=1}) does not satisfy the boundary conditions on a tube, and can only be defined on an infinite plane. 

In conclusion, the family of solutions we have found on the tube satisfies the bound (\ref{BPS}). %

\section{Some explicit examples}

\label{sec:examples}

\begin{figure}[t]
\includegraphics[width=2cm]{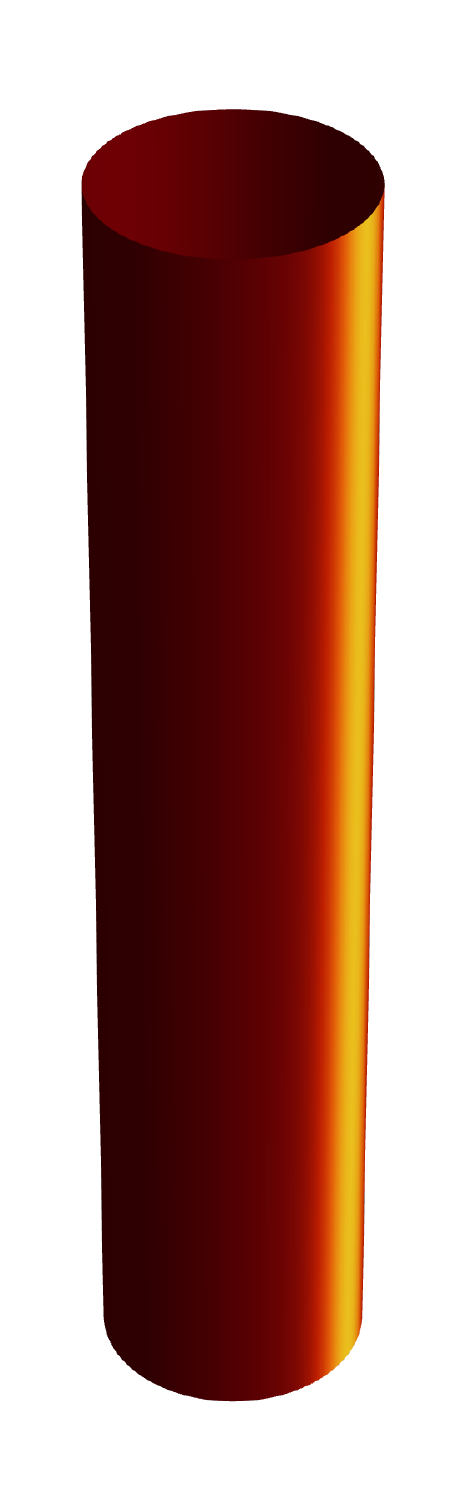}
\includegraphics[width=2cm]{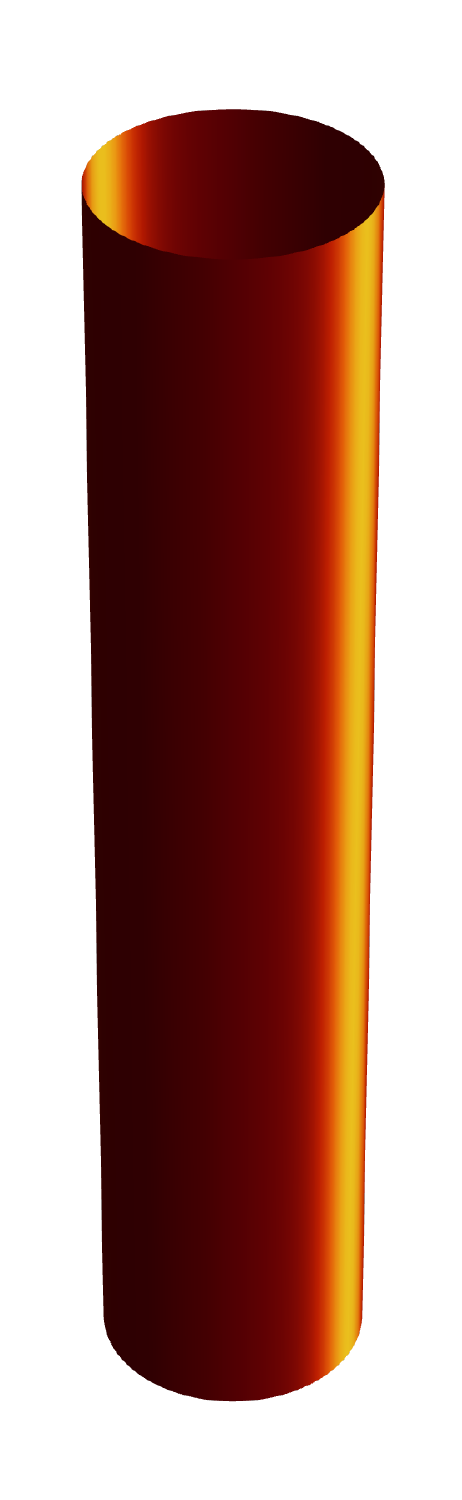}
\includegraphics[width=2cm]{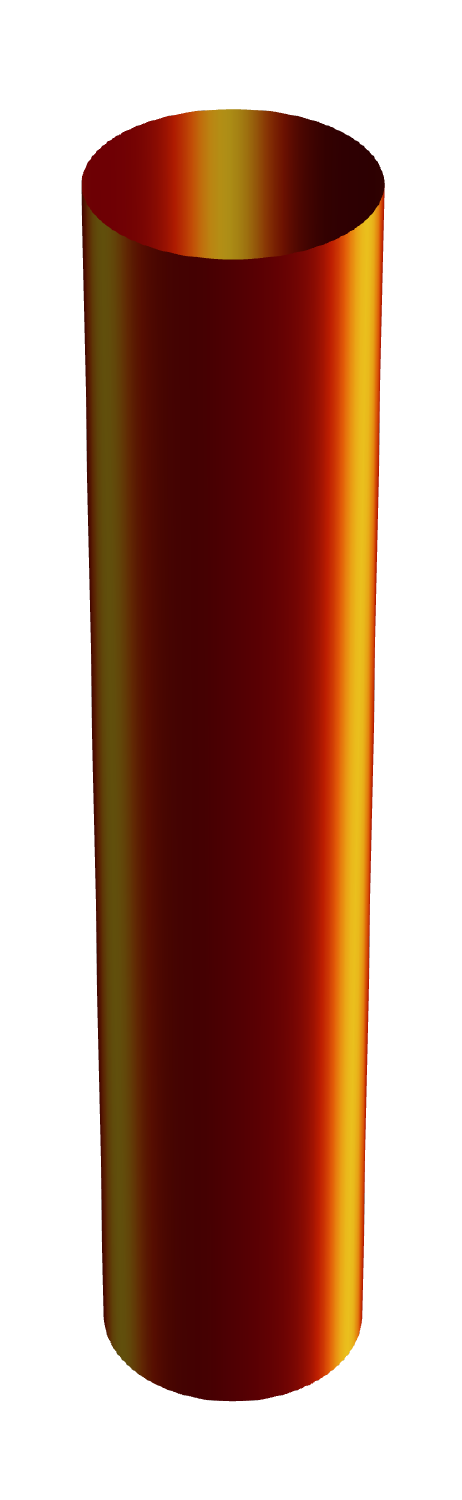}
\\
\includegraphics[width=2cm]{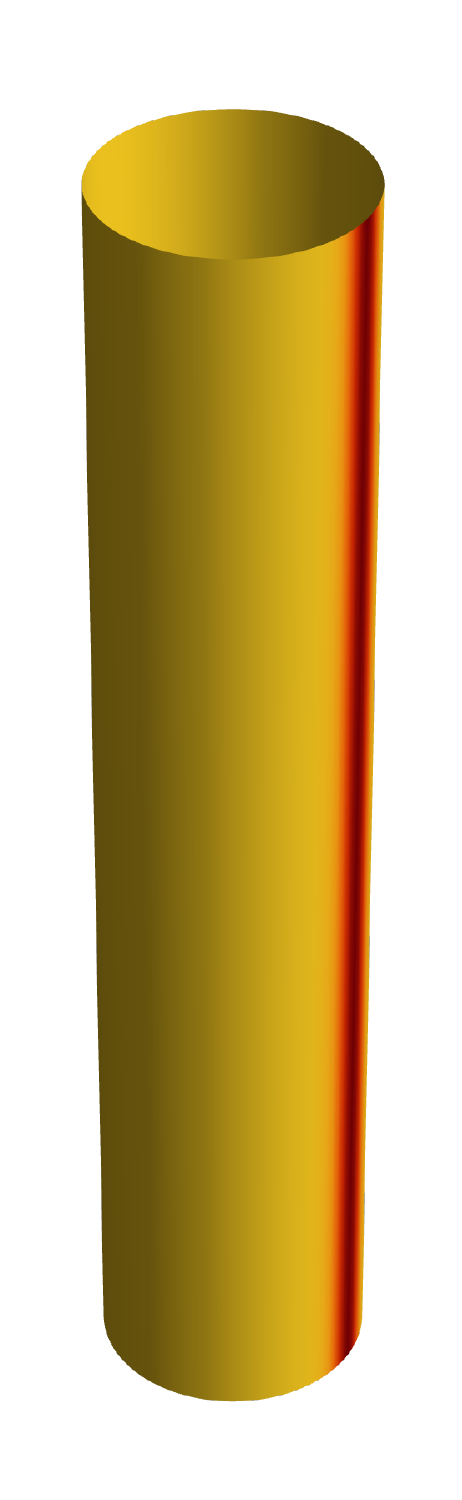}
\includegraphics[width=2cm]{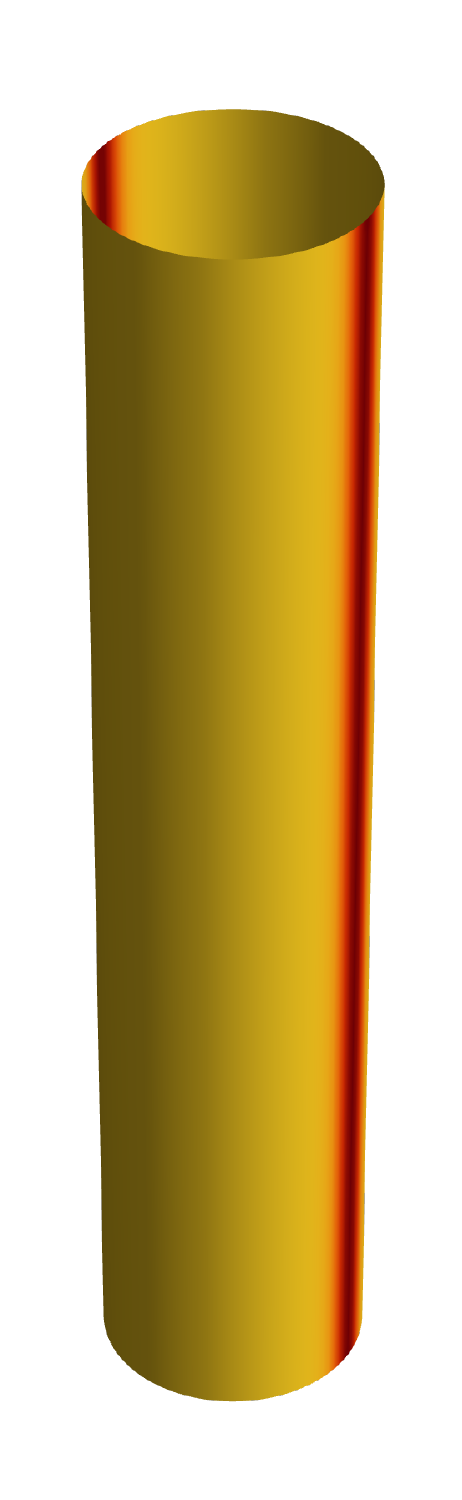}
\includegraphics[width=2cm]{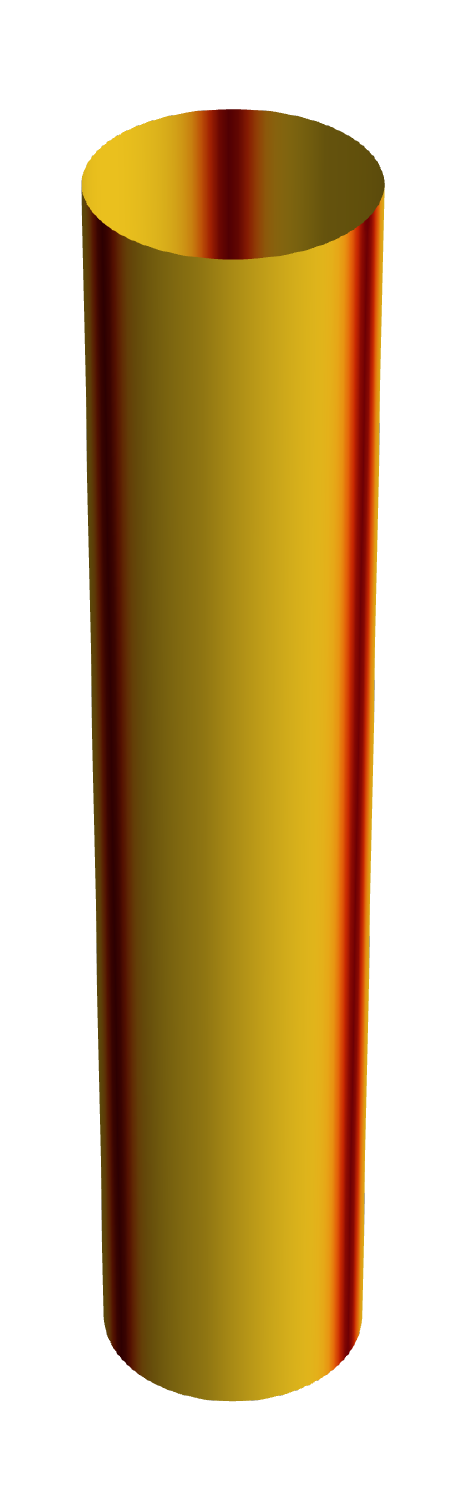}
\\
\includegraphics[width=4cm]{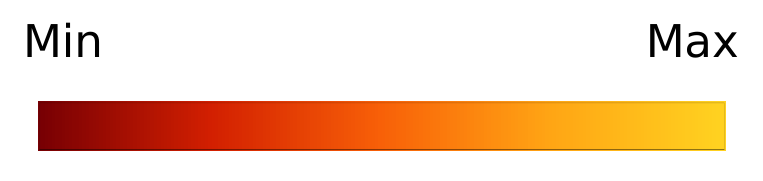}
\caption{Electric current $J_{\pm}$ (top) and energy density $T_{tt}$ (bottom) for $n=1,2,3$ from left to right, with non-vanishing $b_0^\varphi, \tilde b_{00}^\lambda$ and $a_0^F$. It is evident that, as expected, the number $n$ controls the number of maxima around the cylinder.}
\label{figura-n}
\end{figure}

\begin{figure}[ht]
\includegraphics[width=2cm]{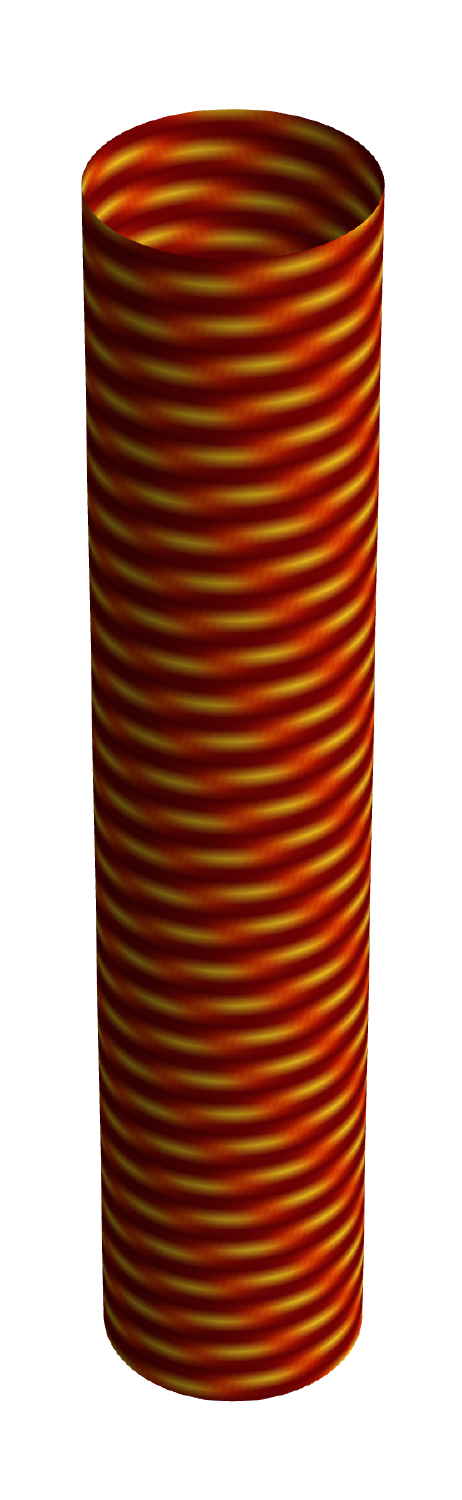}
\includegraphics[width=2cm]{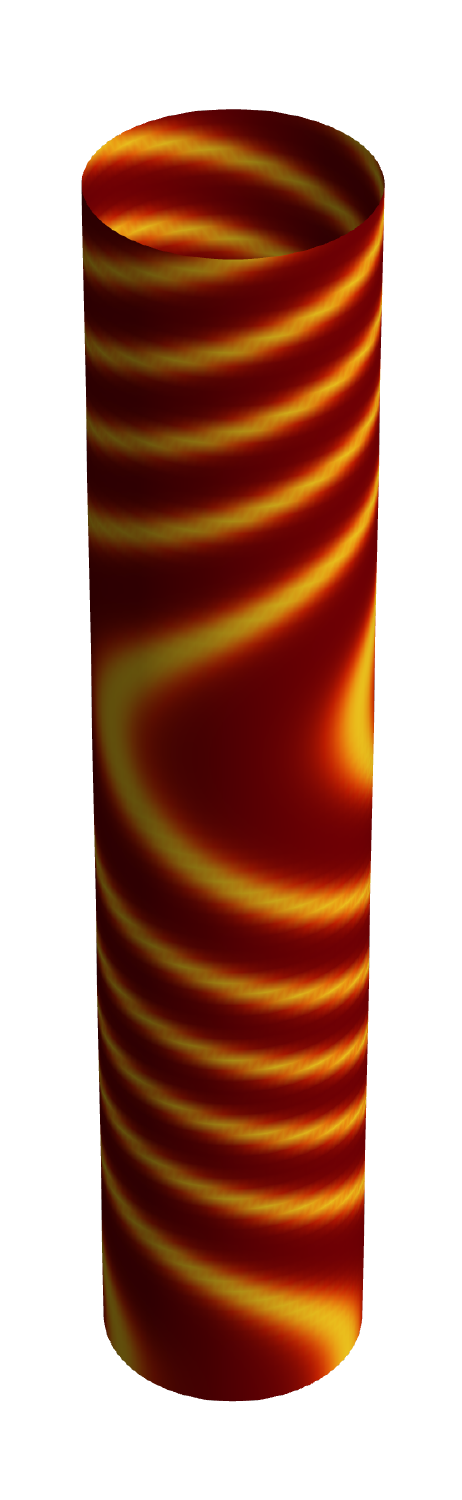}
\includegraphics[width=2cm]{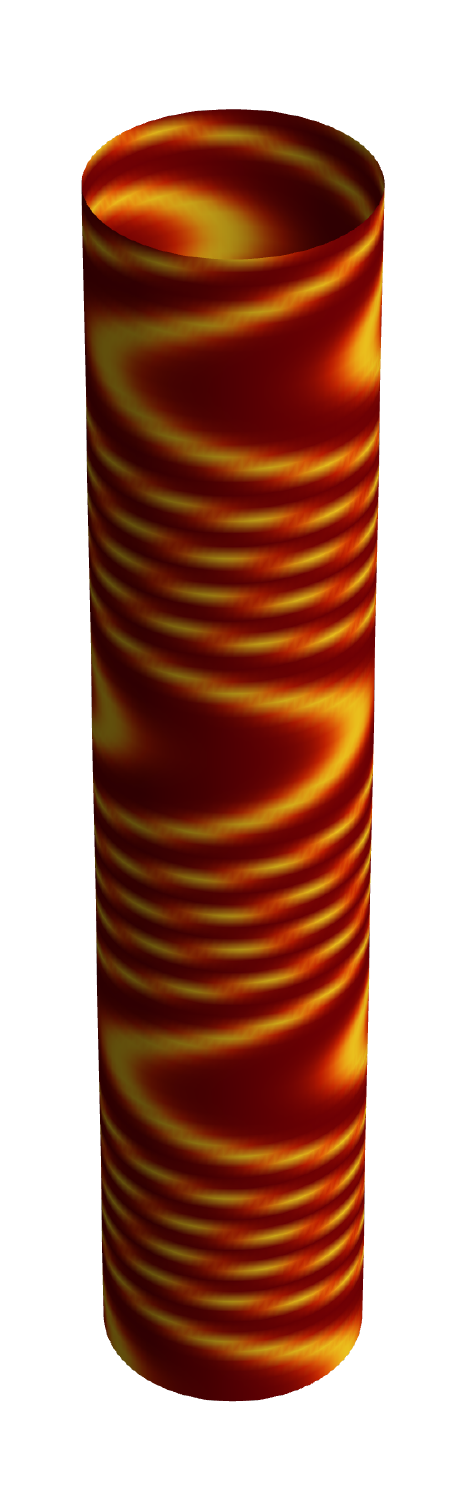}
\\
\includegraphics[width=2cm]{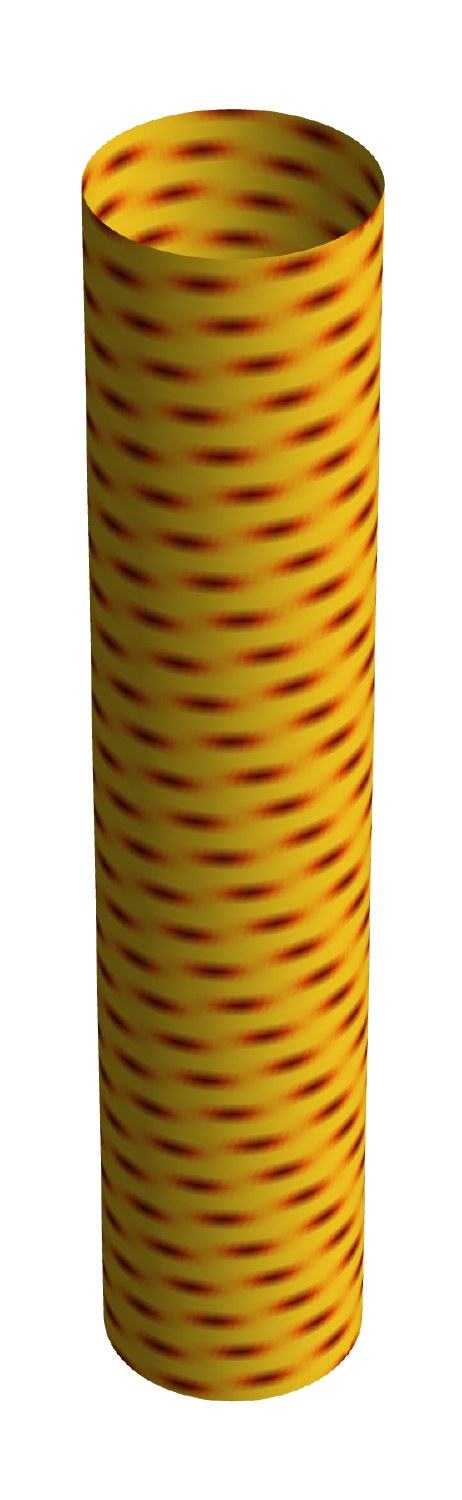}
\includegraphics[width=2cm]{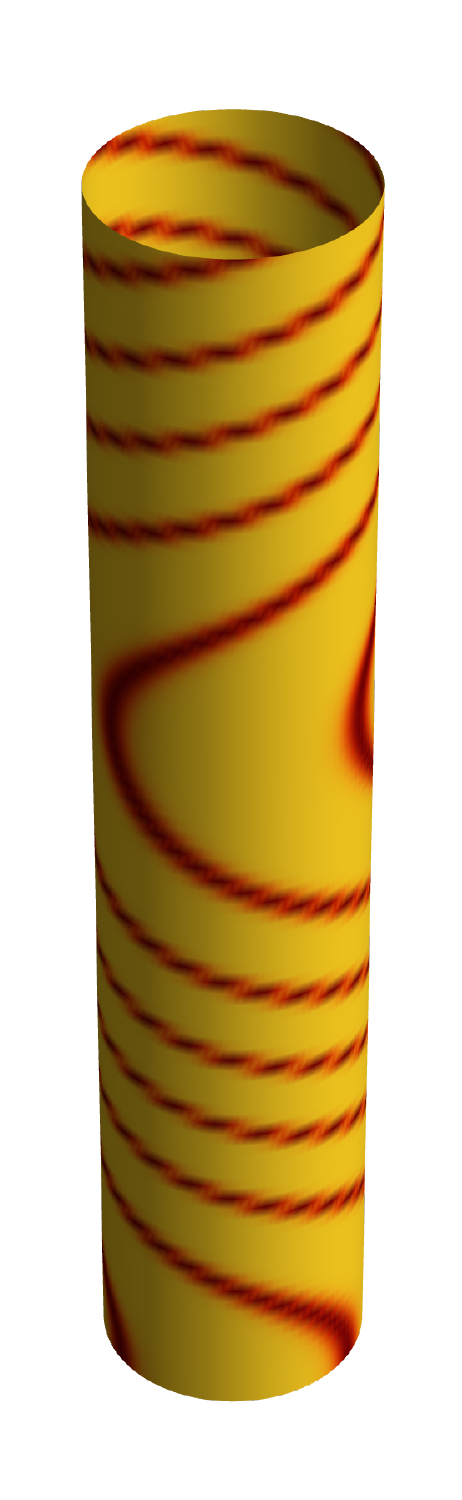}
\includegraphics[width=2cm]{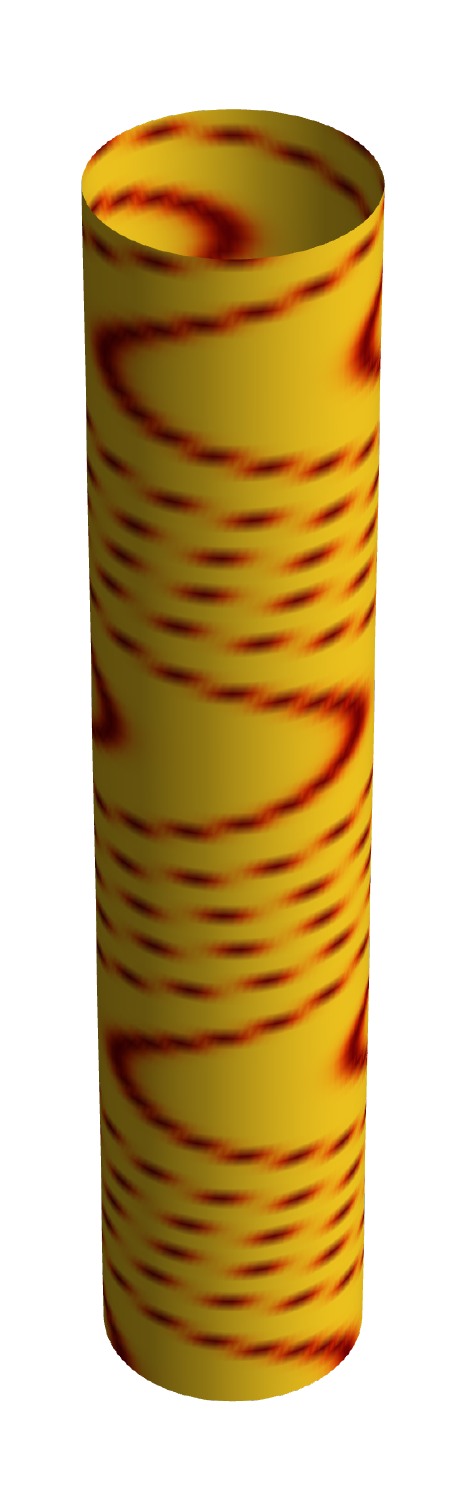}
\\
\includegraphics[width=4cm]{legend.pdf}
\caption{Electric current $J_{\pm}$ (top) and energy density $T_{tt}$ (bottom) for non-vanishing $a^\varphi_0$, $b^\varphi_1$ and $b^\varphi_2$ from left to right, with $n=3$ and non-vanishing $\tilde b_{00}^\lambda$ and $a_0^F$. We see that $a_0^\varphi$ regulates the winding of the level curves around the cylinder, while $k^\varphi$ is counting their oscillations along $z$.}
\label{figura-kphi}
\end{figure}

\begin{figure}[ht]
\includegraphics[width=2cm]{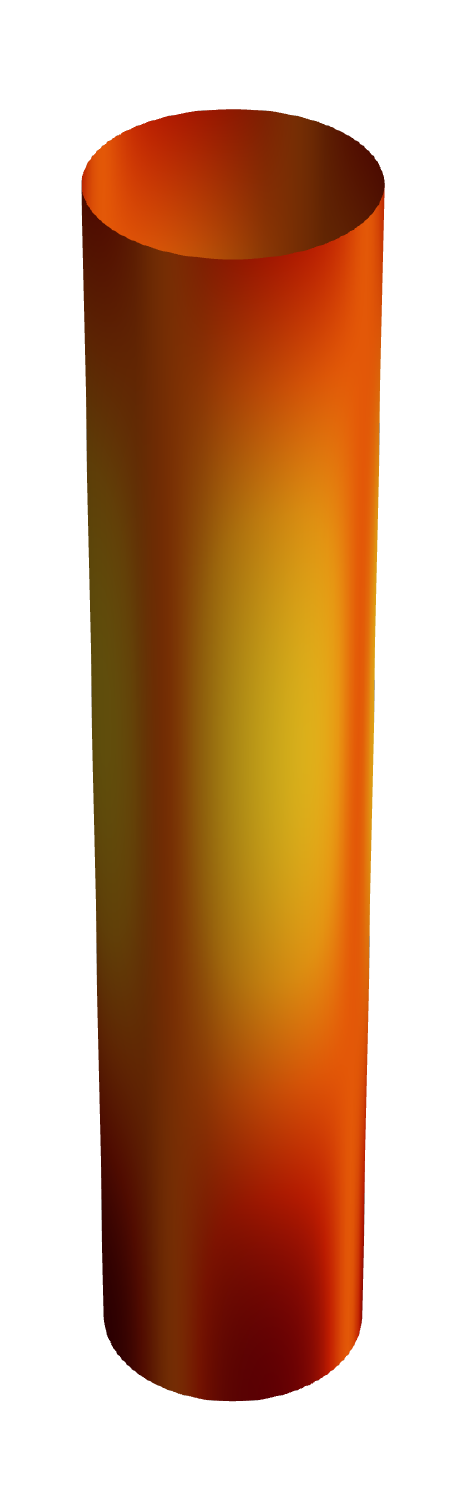}
\includegraphics[width=2cm]{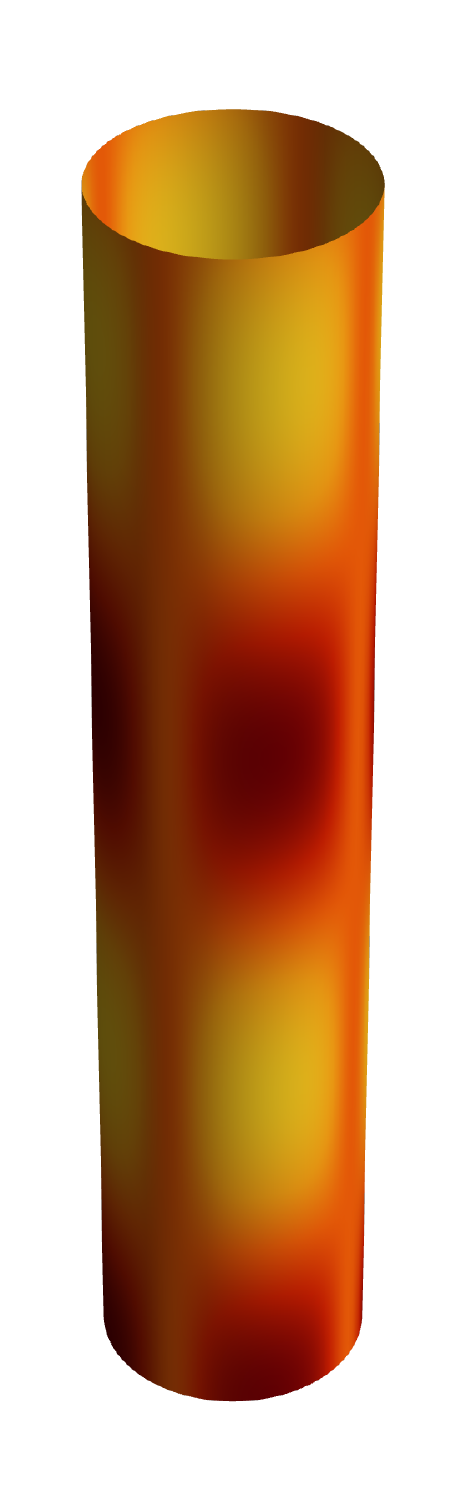}
\includegraphics[width=2cm]{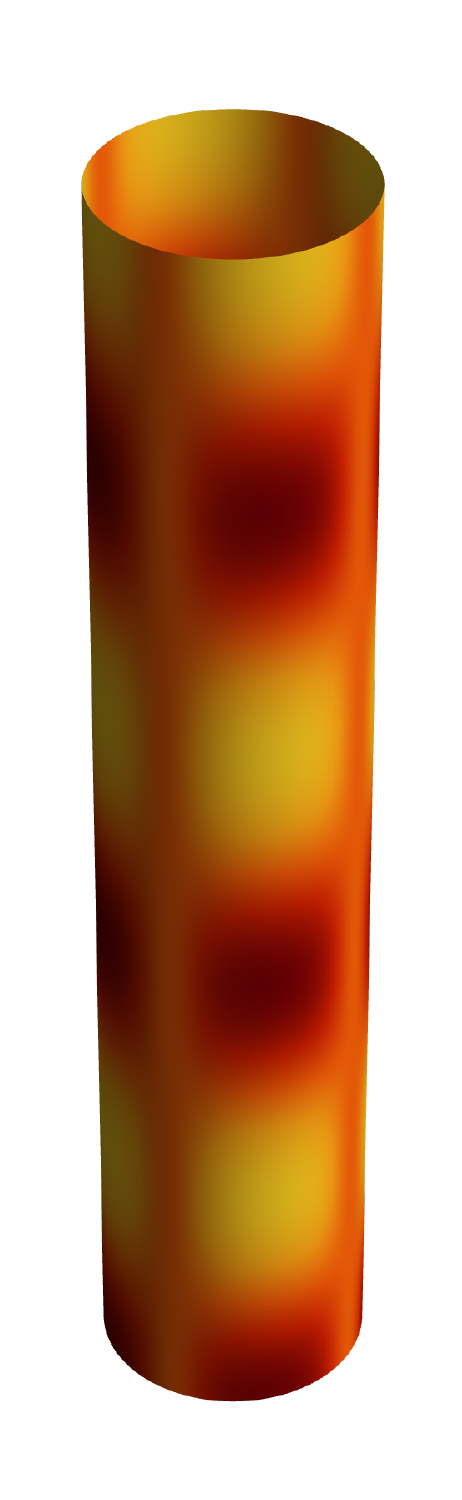}
\\
\includegraphics[width=2cm]{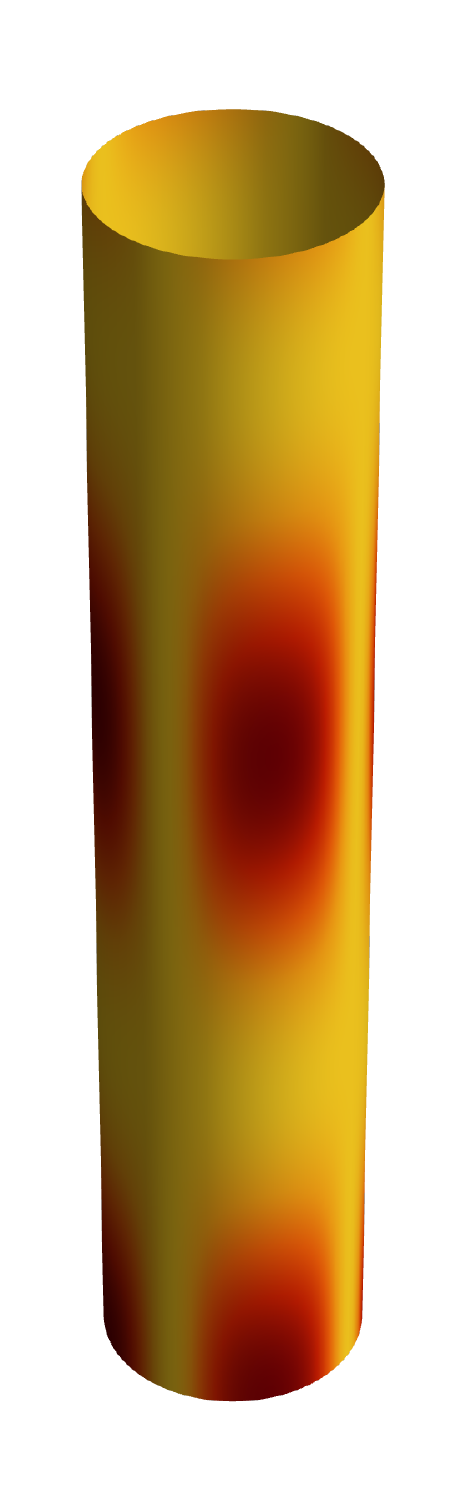}
\includegraphics[width=2cm]{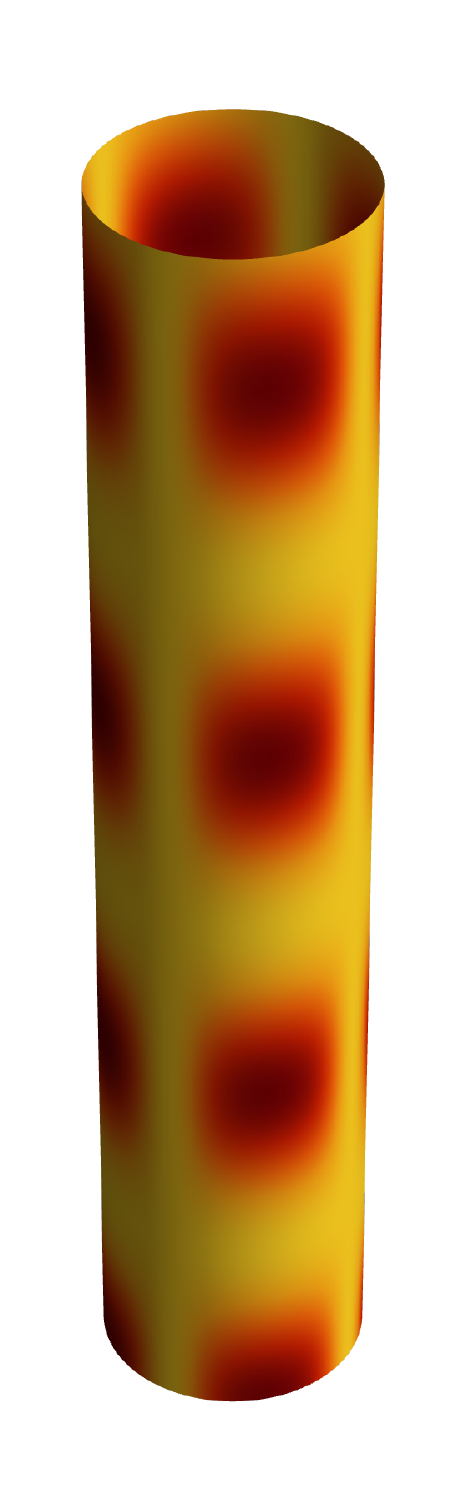}
\includegraphics[width=2cm]{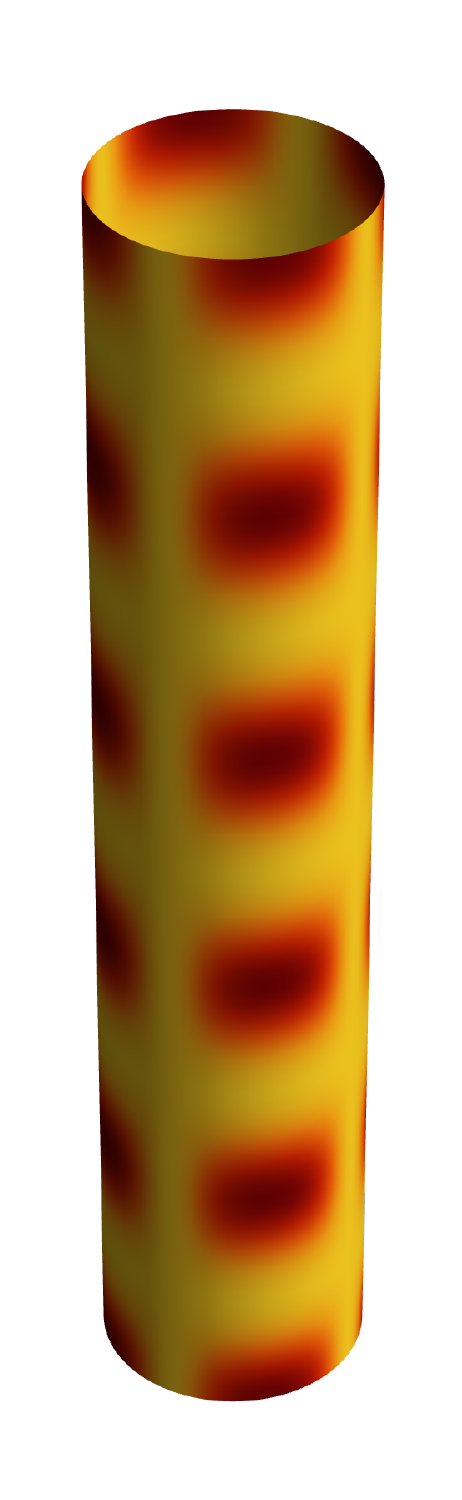}
\\
\includegraphics[width=4cm]{legend.pdf}
\caption{Electric current $J_{\pm}$ (top) and energy density $T_{tt}$ (bottom) for non-vanishing $\tilde b^\lambda_{00}$, $\tilde b^\lambda_{10}$ and $\tilde b^\lambda_{20}$ from left to right, with $n=4$ and non-vanishing $b_{0}^\varphi$ and $a_0^F$. Notice that ${k}^\lambda$ counts the number of maxima along $z$.}´
\label{figura-klambda}
\end{figure}

\begin{figure}[ht]
\includegraphics[width=2cm]{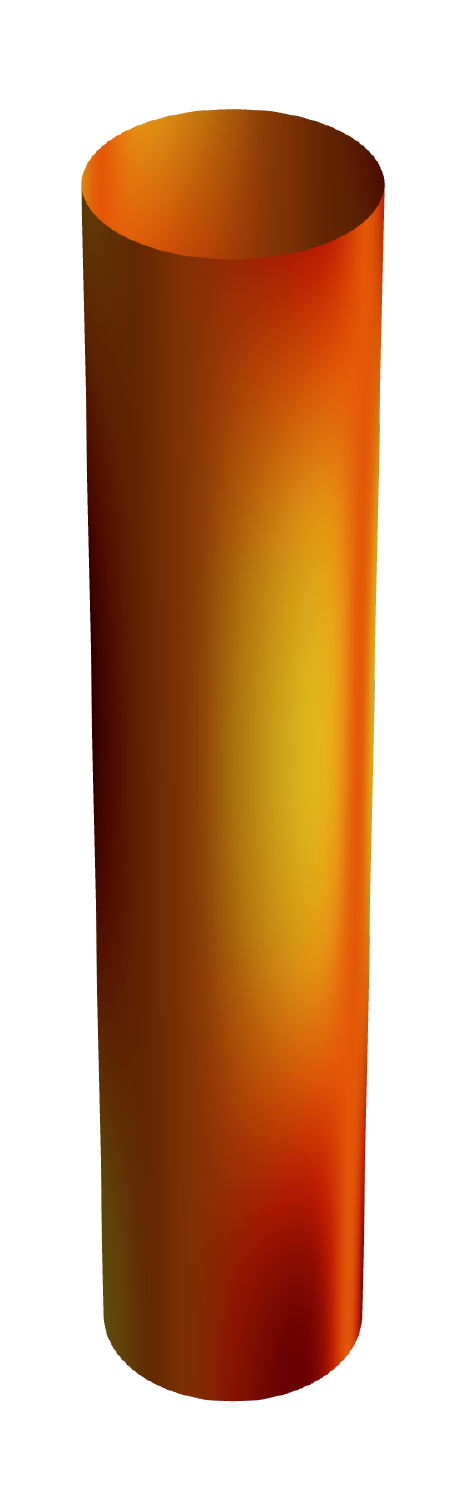}
\includegraphics[width=2cm]{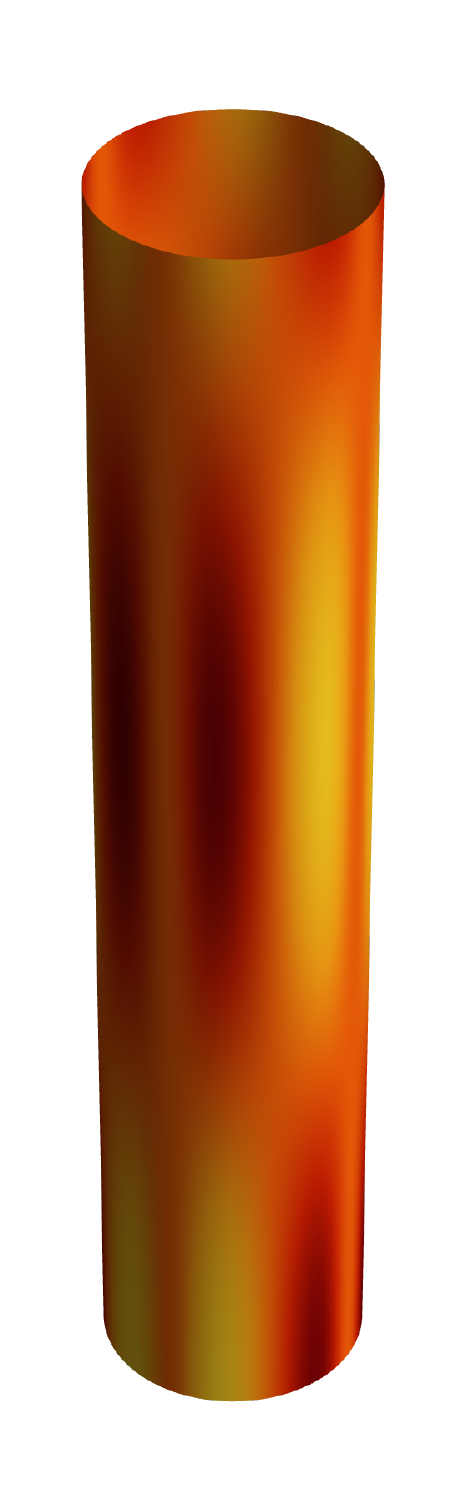}
\includegraphics[width=2cm]{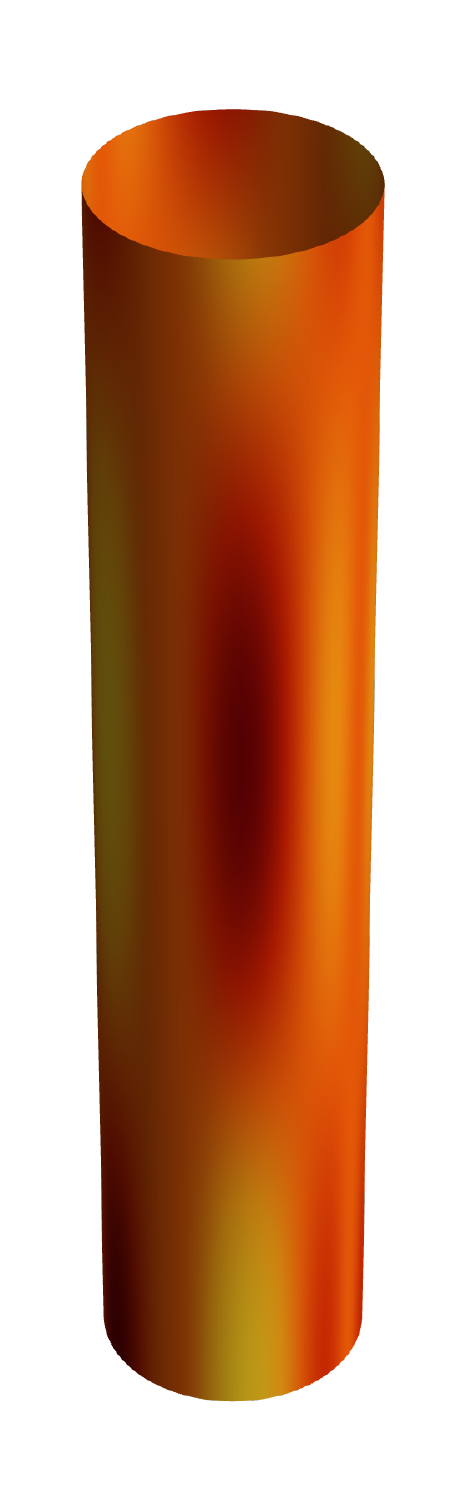}
\\
\includegraphics[width=2cm]{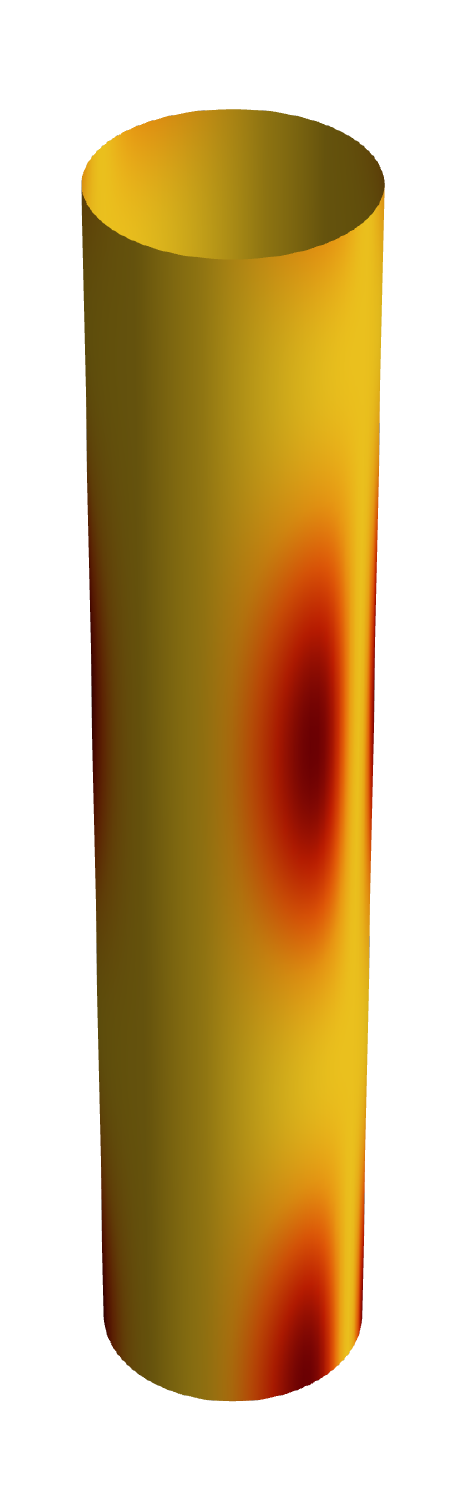}
\includegraphics[width=2cm]{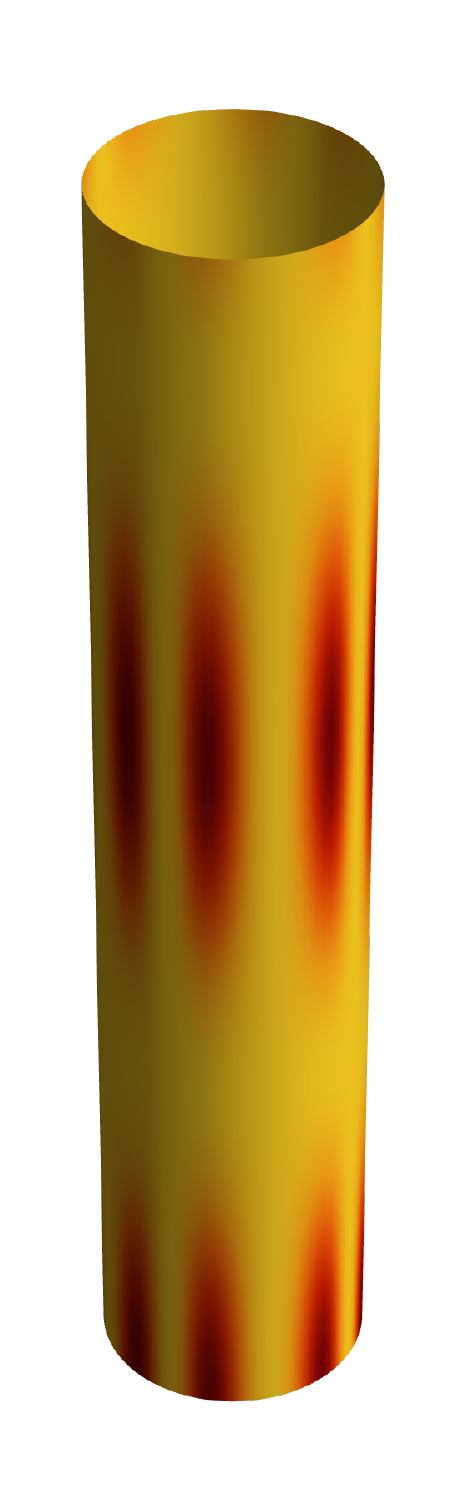}
\includegraphics[width=2cm]{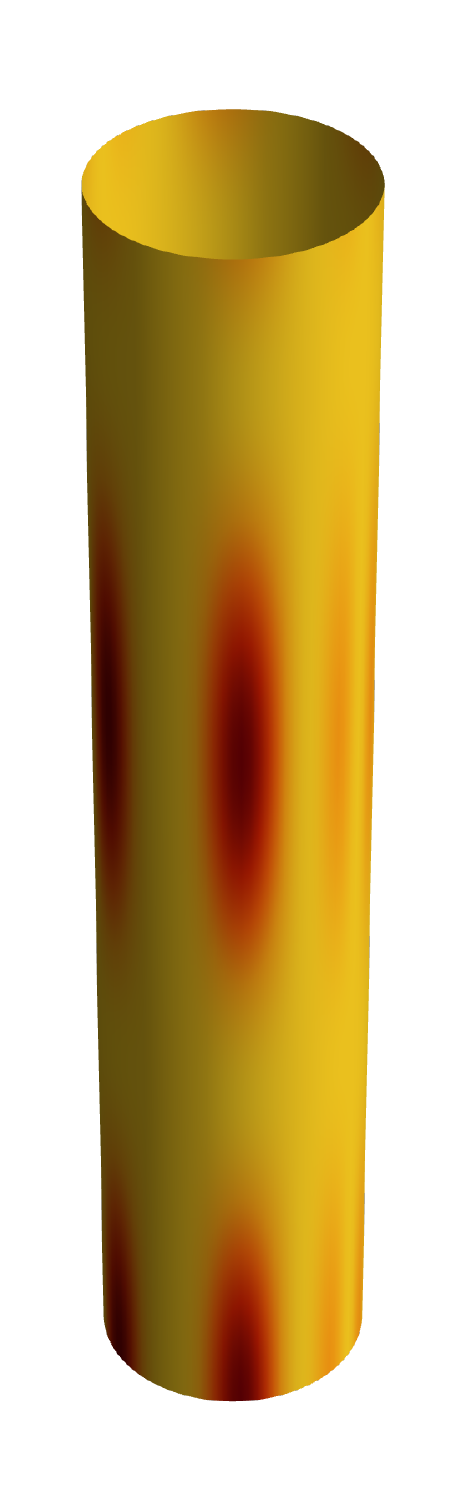}
\\
\includegraphics[width=4cm]{legend.pdf}
\caption{Electric current $J_{\pm}$ (top) and energy density $T_{tt}$ (bottom) for non-vanishing $\tilde b^\lambda_{11}$, $\tilde b^\lambda_{12}$ and $\tilde b^\lambda_{13}$ from left to right, with $n=4$ and non-vanishing $b_{0}^\varphi$ and $a_0^F$. Here $l$ combines with $n$ to control the number of maxima around the cylinder.}´
\label{figura-l}
\end{figure}

\begin{figure}[!ht]
\includegraphics[width=2cm]{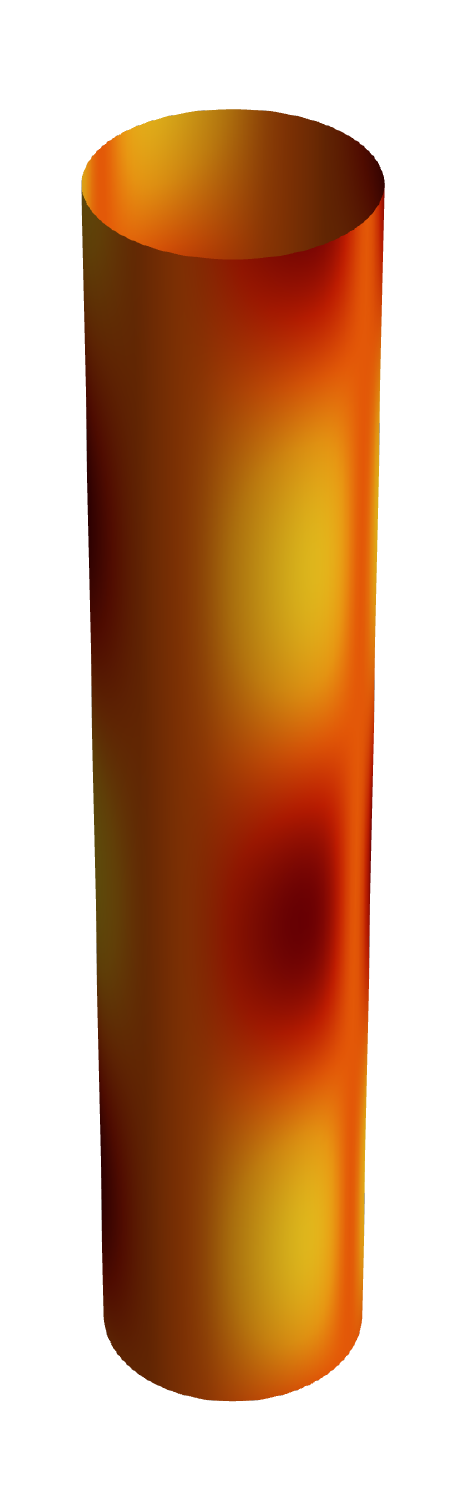}
\includegraphics[width=2cm]{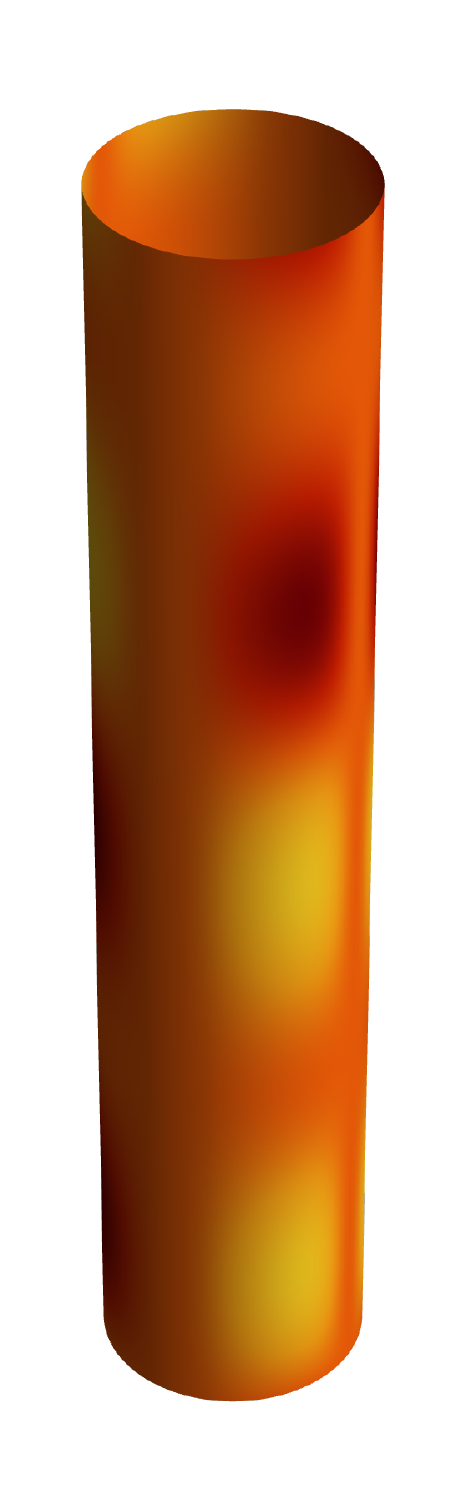}
\includegraphics[width=2cm]{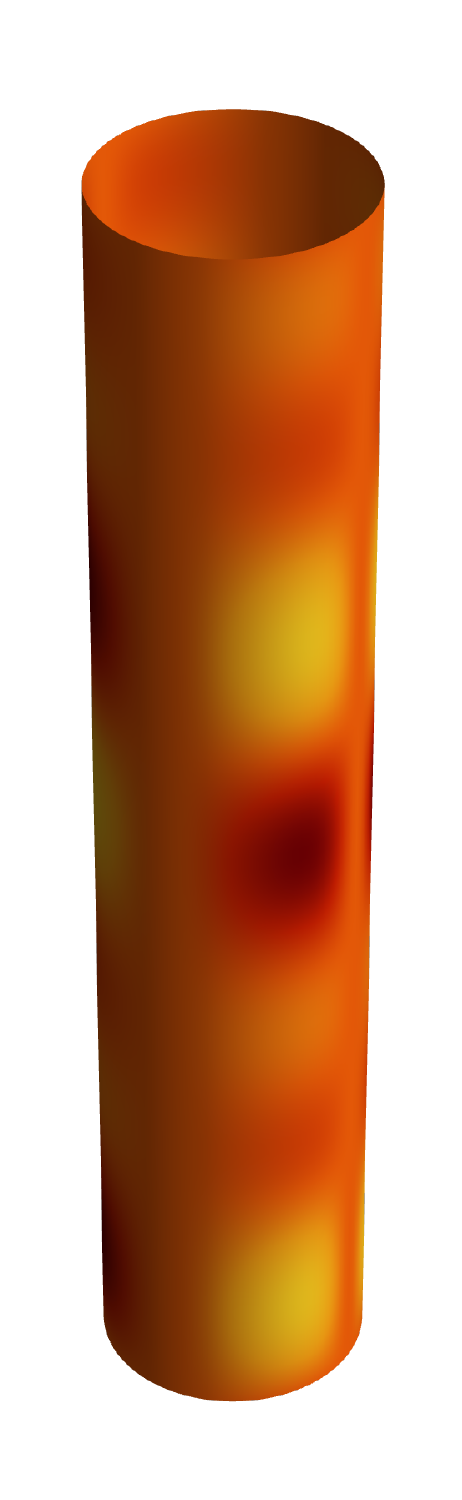}
\\
\includegraphics[width=2cm]{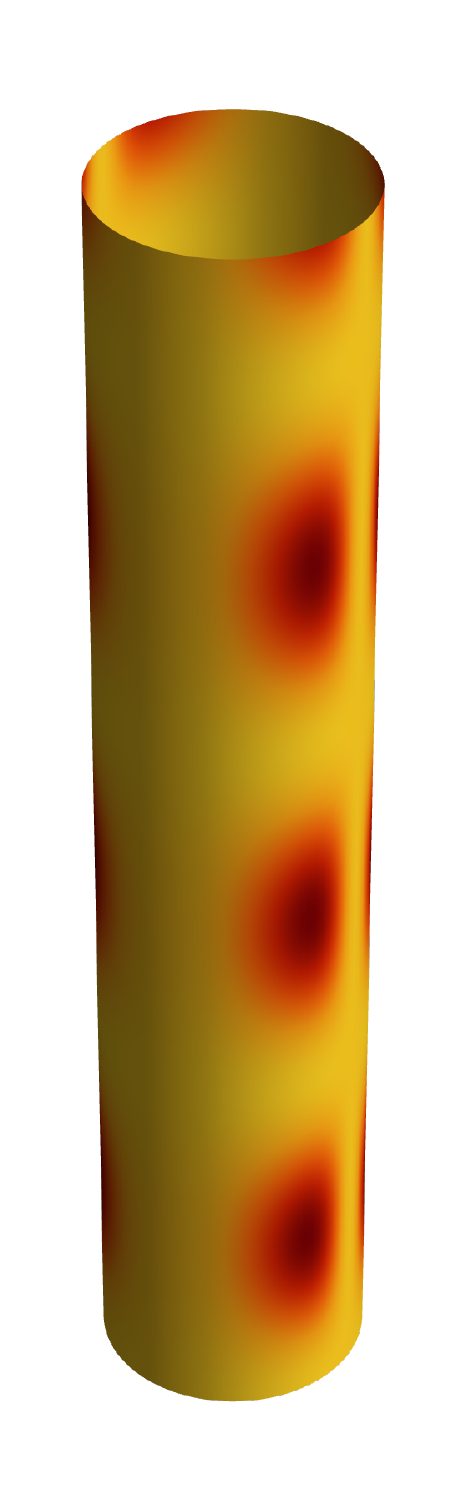}
\includegraphics[width=2cm]{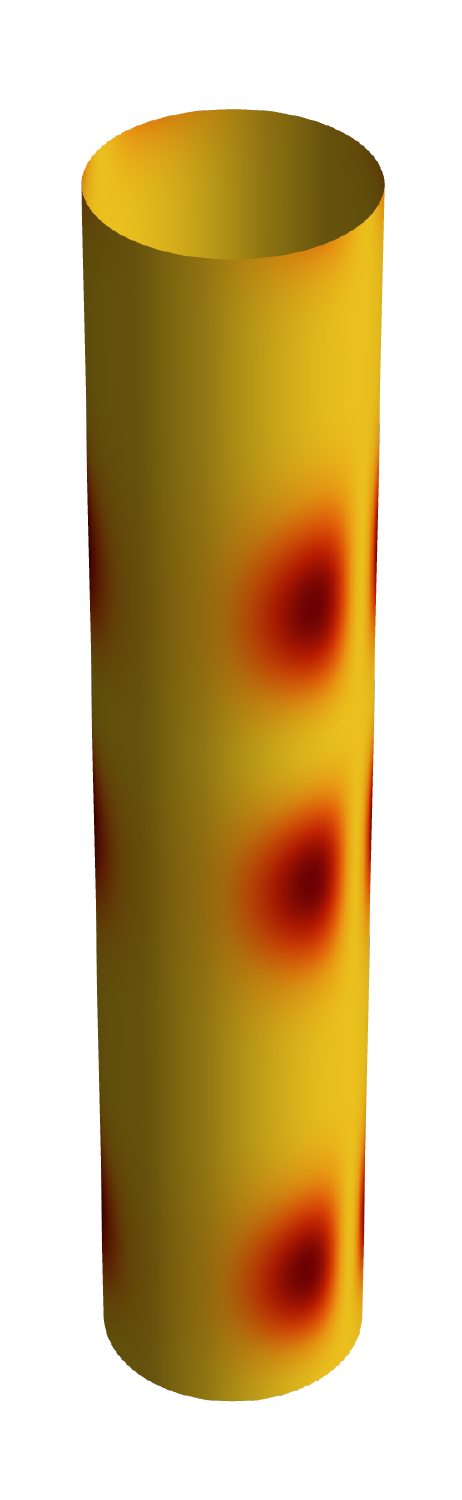}
\includegraphics[width=2cm]{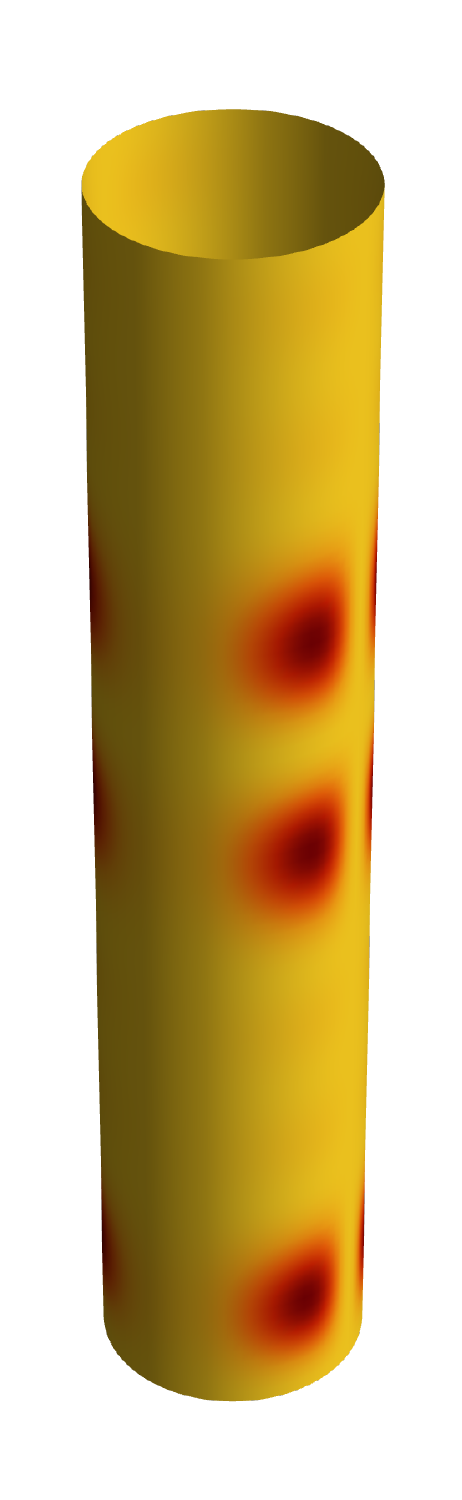}
\\
\includegraphics[width=2cm]{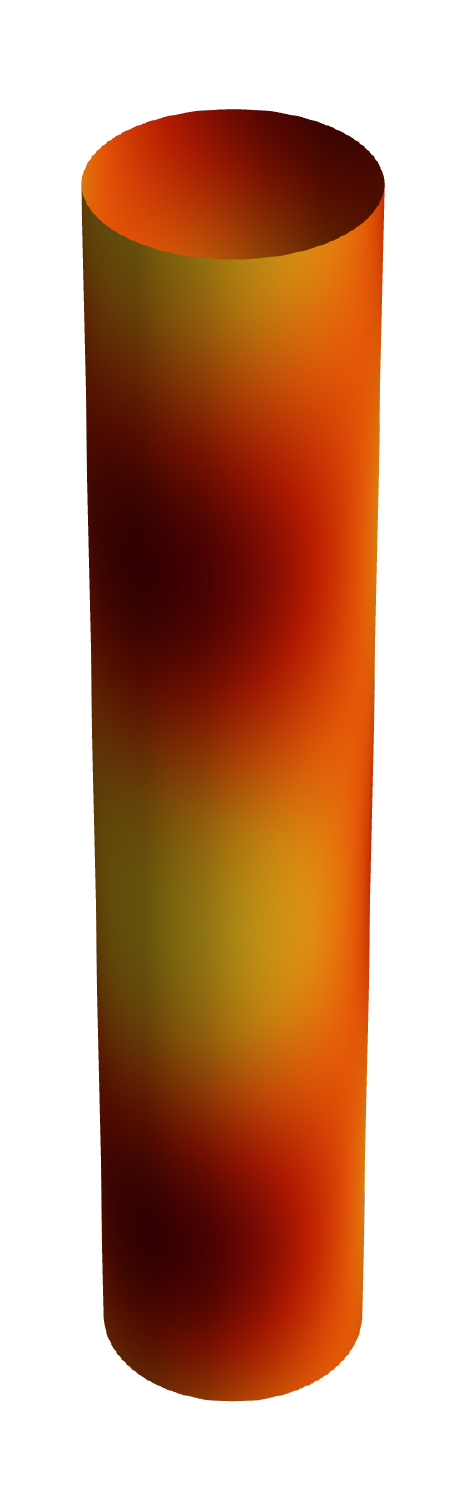}
\includegraphics[width=2cm]{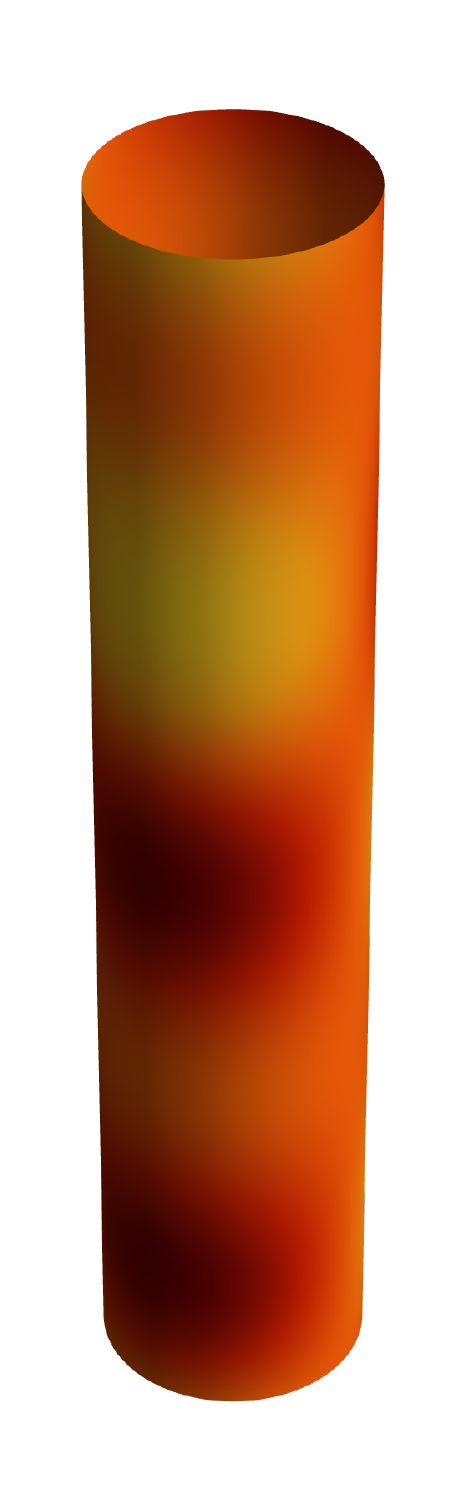}
\includegraphics[width=2cm]{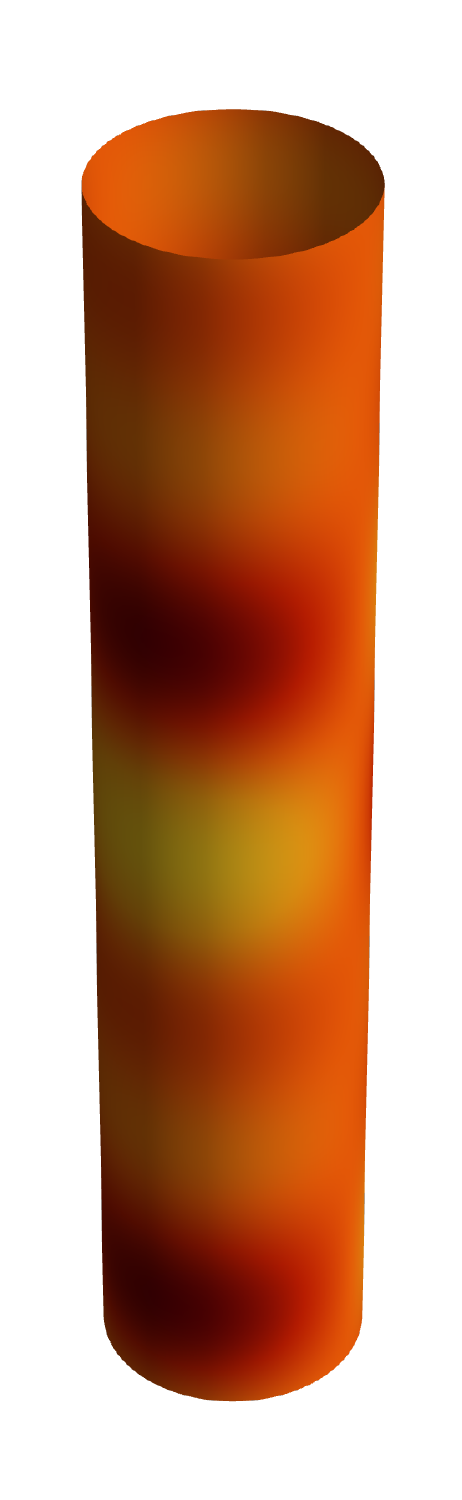}
\\
\includegraphics[width=4cm]{legend.pdf}
\caption{Electric current $J_{\pm}$ (top), energy density $T_{tt}$ (center) and electromagnetic field $E_{\varphi}=\mp B$ (bottom) for non-vanishing $b^F_{1}$, $b^F_{2}$ and $b^F_{3}$ from left to right, with $n=4$ and non-vanishing $b_{0}^\varphi$ and $\tilde b_{11}^\lambda$. We see that $k^F$ combines with $K^\lambda$ to control the number of maxima along the cylinder.}´
\label{figura-kF}
\end{figure}

In order to get some insight on the behavior of the solutions, we need to
specify an explicit form for the arbitrary chiral functions $F(z_\pm)$ and $%
\varphi_0(z_\pm)$, and for the function $\lambda(z_\pm,\varphi)$. We can
write Fourier decompositions for all of them, as 
\small 
\begin{eqnarray}  \label{expansion1}
&& \!\!\!\!\!\!\! F(z_\pm)= a_0^F z_\pm +\!\!\sum_{{k^F}=0}\!a_{k}^F\sin \left(\frac{%
2\pi k^F}{L_z} z_\pm \!\right) +b_{k}^F\cos \left(\frac{2\pi k^F}{L_z}
z_\pm\! \right) ,  \nonumber \\
&&\!\!\!\!\!\!\! \varphi_0(z_\pm)= a_0^\varphi z_\pm\!
+\!\!\sum_{k^\varphi=0}\!a_{k}^\varphi\sin \left( \frac{2\pi k^\varphi}{L_z}z_\pm \right)
+b_{k}^\varphi\cos \left( \frac{2\pi k^\varphi}{L_z}z_\pm\! \right) , 
\nonumber \\
&&\!\!\!\!\!\!\! \lambda(z_\pm)= a_0^\lambda z_\pm\!
+\!\!\sum_{k^\lambda=0}\!a_{k}^\lambda\sin \left( \frac{2\pi k^\lambda}{L_z}z_\pm \right)
+b_{k}^\lambda\cos \left( \frac{2\pi k^\lambda}{L_z}z_\pm\! \right)  , 
\nonumber
\end{eqnarray}%
\normalsize 
where the coefficients $a_k^\lambda$ and $b_k^\lambda$ are
periodic functions of $\varphi$, and can be decomposed according to 
\begin{eqnarray}
&&\!\!\!\!\!\!\! a_k^\lambda(\varphi)= \sum_{l=0}a^\lambda_{kl}\sin \left( 
\frac{2\pi l}{L_\varphi}\varphi\right) +\tilde a^\lambda_{kl}\cos \left( 
\frac{2\pi l}{L_\varphi}\varphi\! \right) \, ,  \nonumber \\
&&\!\!\!\!\!\!\! b_k^\lambda(\varphi)= \sum_{l=0}b^\lambda_{kl}\sin \left( 
\frac{2\pi l}{L_\varphi}\varphi\right) +\tilde b^\lambda_{kl}\cos \left( 
\frac{2\pi l}{L_\varphi}\varphi\! \right) \, .  \nonumber
\end{eqnarray}%
With this expressions, we can plot the profiles of the observable functions, namely the electric current $J_\pm$, the energy density $T_{tt}$, and the electromagnetic field $E=\mp B$, for some simple examples, see Figs.~\ref{figura-n} to \ref{figura-kF}. 

In Fig.~\ref{figura-n} we plot some simple configurations with growing values of $n$, which implies a growing number of maxima of the functions around the cylinder. In Fig.~\ref{figura-kphi} we draw some solutions with different values of $k^\varphi$, the linear mode controlling the winding of the level curves around the cylinder, the higher modes counting their oscillations along it. Fig.~\ref{figura-klambda} shows configurations with different values of $k^\lambda$, which controls the number of maxima of the functions along the cylinder. Fig.~\ref{figura-l} shows the some profiles with fixed $k^\lambda$ for different values of $l$, which combines with $n$ to tweak the number of maxima around the cylinder. Finally Fig.~\ref{figura-kF} contains profiles with different values of $k^F$, contributing to the number of maxima of the observable functions along the cylinder.

\newpage
\section{\protect\normalsize Stability}
\label{sec:stability} 

We proved in Section \ref{sec:flux} that our solutions are characterized by a topological charge $Q=2\pi n$, in
which $n$ represent the winding around the cylinder of the phase of the Higgs field. This feature is a proxy for
the overall stability of the configuration, since a finite-energy deformation cannot change the winding number.
Then, if an instability exists, it must drive the solution into a different one with the same winding number,
as for example the Abrikosov-Nielsen-Olesen vortex. 

A complete perturbative analysis of the obtained solutions 
is beyond the scope of the present paper, since it would involve five
coupled linear partial differential equations on a non-trivial background. 
Nevertheless, in this section we analyze a special type of perturbations 
which have both interesting physical meaning and allow some analytic
control: those that preserve the decoupling properties of our Ansatz. Such
perturbations are defined by the following small deformations of our Ansatz
functions 
\begin{eqnarray}
F(z_{\pm })&\rightarrow& F(z_{\pm })+\varepsilon \,\delta F(z_{\pm })\ , \\
h(\varphi)&\rightarrow& h(\varphi)+\varepsilon\, \delta h(\varphi)\ ,
\label{pertype1}
\end{eqnarray}
where $\varepsilon$ is a small dimensionless parameter. These perturbations
are very likely to be the smallest energy perturbations of the present exact
solutions. The reason is that it takes ``a little effort" to perform an
angular deformation of $h$ (see, for instance, the discussion of the
hedgehog ansatz\cite{shifman1, shifman2}) as compared to deformations on $h$
depending also on the other coordinates. For this reason the linear operator
that determines the spectrum of these perturbations plays a very important
role. 

The linearized field equations are obtained expanding to the
first nontrivial order in $\varepsilon$. They read 
\begin{eqnarray}
\partial_{\pm}F\,\partial_{\mp}\delta F&=&0\ ,  \label{F1yF2} \\
-\partial_\varphi^2\delta h+\gamma (3h^{2}-\nu ^{2}) \delta h &=&0\ ,
\end{eqnarray}%
where $h$ and $F$ are the background solutions. 

As far as the equation for the perturbation $\delta F$ is
concerned, since $F$ only depends on one of the two light-cone variables $z_\pm$ we
get that the solution for $\delta F$ must depend on the same light-cone
variable 
\begin{equation}
\delta F=\delta F\left( z_{\pm}\right) \ .
\end{equation}
Thus, imposing reasonable boundary conditions in the $z$-direction, 
one gets a Fourier expansion for $\delta F$ with real frequencies, so that
the perturbation is always well-behaved. 

Regarding the perturbation $\delta h$, if we take 
\begin{equation}
\delta h=\partial _{\varphi }h \ ,
\end{equation}
where $h$ is the background solution satisfying Eq. (\ref{QuOsc}), then $%
\delta h$ satisfies identically the corresponding linearized field equation
in Eq. (\ref{F1yF2}). Moreover, from Eq. (\ref{minustphiphi}) it is clear
that we can choose the integration constant $m$ close enough to one, in such
a way that $\partial _{\varphi }h$ never changes sign: this implies that $%
\delta h$ has no node. In other words, we have a nodeless zero mode of the
linearized field equations. Standard arguments in quantum mechanics then suggest
that all the other eigenvalues 
are positive. 

Although the above arguments are not a complete proof of the
stability, they suggest that the family of analytic solutions constructed
here have interesting physical properties. 

\section{\protect\normalsize Discussion}

\label{sec:discussion} 

We have found exact solutions to $2+1$ scalar electrodynamics, that represent solitonic configurations propagating along a cylinder. The solutions are topological in nature, being indexed by an integer number that counts the winding of the Higgs phase around the cylinder. They are continuously connected to Abrikosov-Nielsen-Olesen vortices. Even if a partial perturbative analysis does not show instabilities, our solutions may in principle relax into vortices via some instability channel we had not considered, or through external dissipation. 

There are both a charge density and an electric current along the cylinder, which are equal (up to a sign) and have arbitrary shapes on the cylinder surface. The current is self-sustained since there is no need of an external field to keep it alive. 
There is no electric current around the cylinder, nor electric field along it. 
As time runs, the profile moves as a whole along the cylinder, at the speed corresponding to that of the light in the model.

We would like to interpret our solutions as long standing excitations on a superconducting nanotube. As these configurations are independent of the value of the coupling, they can describe both type I and type II superconductors. For this interpretation to work, we need that (1) the superconducting condensate has an $s$-wave symmetry\cite{swave1, swave2, swave3}, (2) the matter dynamics can be considered relativistic\cite{holandos, hirsch, grigorishin}, and (3) our two-dimensional fields have to be embedded into a three dimensional setup (this can be done by solving Maxwell's equation in vacuum and then imposing at the tube radius a suitable set of boundary conditions, that in cylindrical coordinates read $B_r=B$, $E_z=0$, $E_\varphi=\pm F_{\varphi\pm}$). If these hypotheses are fulfilled, the device could in principle be constructed out of twisted bilayer graphene, by compactifying one direction on a certain number of its moir\'e periods of around $\simeq 13$nm.

\section*{\protect\normalsize Aknowledgements}

{\normalsize N.G. is grateful to Mauricio Sturla, Guillermo Silva and Martin
Schvellinger for useful references and insights, and to Universidad de
Concepci\'on and Centro de Estudios Cient\'ificos for hospitality and
support. His work is partially supported by CONICET grants PIP-2017-1109 and
PUE084 ``B\'usqueda de nueva f\'isica'', and UNLP grant PID-X791. A. G has
been funded by FONDECYT Grant No. 1200293. F. C. has been funded by FONDECYT
Grant No. 1200022. J. O. thanks the support of FONDECYT Grant No. 1221504.
A. V. has been funded by FONDECYT post-doctoral Grant No. 3200884. The
Centro de Estudios Cient\'ificos (CECs) is funded by the Chilean Government
through the Centers of Excellence Base Financing Program of ANID. }

\end{document}